\documentstyle[aps,12pt]{revtex}
\begin{document}
\title{Nonlinear dynamics of giant resonances in atomic nuclei} 
\author{ D. Vretenar$^{1,2}$, N. Paar$^{1}$, 
P. Ring$^{2}$, and G.A. Lalazissis$^{2}$}
\address{$^{1}$ Physics Department, Faculty of Science, University of
Zagreb, Croatia\\
$^{2}$ Physik-Department der Technischen Universit\"at M\"unchen,
D-85748 Garching, Germany}
\maketitle
\bigskip
\begin{abstract}
The dynamics of monopole giant resonances in nuclei is analyzed
in the time-dependent relativistic mean-field model. The 
phase spaces of isoscalar and isovector collective 
oscillations are reconstructed from the 
time-series of dynamical variables that characterize the 
proton and neutron density distributions. The analysis
of the resulting recurrence plots and correlation dimensions
indicate regular motion for the isoscalar mode, and chaotic
dynamics for the isovector oscillations. Information-theoretic
functionals identify and quantify the nonlinear dynamics
of giant resonances in quantum systems that have spatial 
as well as temporal structure.  
\end{abstract}
\section{Introduction}

In a recent article~\cite{chaos.97} we have started the analysis
of collective nonlinear dynamics in atomic nuclei in the
framework of time-dependent relativistic mean field theory.
Atomic nuclei provide excellent examples of quantum systems
in which the transition from regular to chaotic dynamics 
can be observed in experiments, and studied with a variety 
of sophisticated theoretical models. Signatures of chaotic
dynamics have been observed in correlations of nuclear 
level distributions, and in the microscopic and collective 
motion of the nuclear many-body system~\cite{Zel.96}.
Especially interesting in this respect are giant resonances:
highly collective nuclear excitations whose properties,
excitation energies and widths, nevertheless reflect the 
underlying microscopic dynamics. Theoretical studies predict
that regular collective modes coexists with chaotic single-nucleon
motion: the adiabatic mean-field created by the nucleons averages
out the random components of their motion. In Ref.~\cite{chaos.97} 
we have studied the dynamics of the most simple giant resonances:
isoscalar and isovector collective monopole oscillations. In these
resonances only spatial degrees of freedom are excited and 
the motion is spherically symmetric, and therefore relatively 
simple for numerical integration.  We have analyzed 
time-dependent and self-consistent calculations that reproduce
the experimental data on monopole giant resonances in 
spherical nuclei. In the microscopic mean-field model,
self-consistent solutions for ground-states provide initial
conditions, and fully time-dependent calculations are 
performed for the single-nucleon dynamics. Due to  
the self-consistent time-evolution,  
the nuclear model system is intrinsically non-linear. 
In particular, we have studied the difference in the 
dynamics of isoscalar and isovector collective modes. 
Time-series, Fourier power spectra, 
Poincar\' e sections, autocorrelation functions, and
Lyapunov exponents have been used to characterize the 
nonlinear system and to identify chaotic oscillations.
It has been shown that the oscillations of the collective
coordinate can be characterized as regular for the 
isoscalar mode, and that they become chaotic when
initial conditions correspond to the isovector mode. 

In the present article we continue our investigation
of giant monopole resonances. There are two main
objectives. Firstly, the analysis of phase spaces 
reconstructed from time-series of collective dynamical
variables that characterize the isoscalar and 
isovector oscillations. The structure of recurrence plots 
and the correlation dimensions should provide additional 
information about the chaotic regime of collective motion.
The second objective is to study possible applications 
of modern information-theoretic techniques to the 
nuclear many-body dynamics (analysis of information entropies
and evaluation of mutual information functions). This is 
especially important since nuclei are quantum objects of 
finite size, and therefore display interesting spatial 
as well as temporal behavior. We will discuss several 
information functionals that could be used to examine 
the influence of the finite spatial extension of nucleon
densities on the nonlinear dynamics of collective excitations.  

The article is organized as follows. In Sec.~2 we describe
the time-dependent relativistic mean-field model. The time-series
analysis and the reconstruction of phase spaces are performed  
in Sec.~3. The use of information-theoretic functionals 
in the description of nonlinear dynamics is discussed in Sec.~4.
A summary of our results is presented in Sec.~5.

\section{Time-dependent mean-field model}

Theoretical models based on quantum hadrodynamics~\cite{SW.86,SW.97}
have been remarkably successful in the description of many physical
phenomena in atomic nuclei, nuclear matter, neutron stars, 
heavy-ion collisions, electron scattering on nuclei. In particular,
the relativistic mean-field model has been applied in calculations
of properties of ground and excited states, both for spherical
and deformed nuclei (for a recent review see~\cite{Rin.96}). 
In Refs.~\cite{VBR.95,RVP.96,PVR.96,Vre.97} we have used the 
time-dependent version of the relativistic mean-field model 
to describe the dynamics of giant resonances. These collective 
excitations have been observed in nuclei over the whole 
periodic table, and their characteristic properties vary smoothly 
with mass number. Giant resonances can therefore be described 
with models based on the mean-field approximation: 
the two-body nucleon-nucleon correlations are replaced by the 
independent motion of nucleons in the effective one-body 
potential. The nucleons themselves are the sources of the potential.
The self-consistent model of nuclear dynamics is essential for 
a correct description of ground and excited states. 
Self-consistent calculations ensure that the same correlations
that are important for the ground states also determine the 
dynamics of excited states, in this particular case giant resonances.

The details of the time-dependent relativistic mean-field 
model can be found in Refs.~\cite{VBR.95,RVP.96}. In this 
section we include a short outline of the basic properties.
In quantum hadrodynamics
the nucleus is described as a system of Dirac nucleons which
interact through the
exchange of virtual mesons and photons. The model is 
based on the one-boson exchange description of the 
nucleon-nucleon interaction. The Lagrangian
density reads
\begin{eqnarray}
{\cal L}&=&\bar\psi\left(i\gamma\cdot\partial-m\right)\psi
~+~\frac{1}{2}(\partial\sigma)^2-U(\sigma )
\nonumber\\
&&-~\frac{1}{4}\Omega_{\mu\nu}\Omega^{\mu\nu}
+\frac{1}{2}m^2_\omega\omega^2
~-~\frac{1}{4}{\vec{\rm R}}_{\mu\nu}{\vec{\rm R}}^{\mu\nu}
+\frac{1}{2}m^2_\rho\vec\rho^{\,2}
~-~\frac{1}{4}{\rm F}_{\mu\nu}{\rm F}^{\mu\nu}
\nonumber\\
&&-~g_\sigma\bar\psi\sigma\psi~-~
g_\omega\bar\psi\gamma\cdot\omega\psi~-~
g_\rho\bar\psi\gamma\cdot\vec\rho\vec\tau\psi~-~
e\bar\psi\gamma\cdot A \frac{(1-\tau_3)}{2}\psi\;.
\label{lagrangian}
\end{eqnarray}
The Dirac spinor $\psi$ denotes the nucleon with mass $m$.
$m_\sigma$, $m_\omega$, and $m_\rho$ are the masses of the
$\sigma$-meson, the $\omega$-meson, and the $\rho$-meson,
and $g_\sigma$, $g_\omega$, and $g_\rho$ are the
corresponding coupling constants for the mesons to the
nucleon. $U(\sigma)$ denotes the nonlinear $\sigma$
self-interaction
\begin{equation}
U(\sigma)~=~\frac{1}{2}m^2_\sigma\sigma^2+\frac{1}{3}g_2\sigma^3+
\frac{1}{4}g_3\sigma^4,
\label{NL}
\end{equation}
and $\Omega^{\mu\nu}$, $\vec R^{\mu\nu}$, and $F^{\mu\nu}$
are field tensors.

From the Lagrangian density the set of coupled equations of
motion is derived: the Dirac equation for the nucleons
\begin{eqnarray}
i\partial_t\psi_i&=&\left[ {\mathbf{\alpha}}
\left(-i{\mathbf{\nabla}}-g_\omega{\mathbf{\omega}}-
g_\rho\vec\tau\vec{\mathbf{\rho}}
-e\frac{(1-\tau_3)}{2}{\mathbf{ A}}\right)
+\beta(m+g_\sigma \sigma)\right. \nonumber\\
&&\left. +g_\omega \omega_0+g_\rho\vec\tau\vec\rho_0
+e\frac{(1-\tau_3)}{2} A_0
\right]\psi_i,
\label{dirac}
\end{eqnarray}
and the Klein-Gordon equations for the meson fields
\begin{eqnarray}
\left(\partial_t^2-\Delta+m^2_\sigma\right)\sigma&=&
-g_\sigma\rho_s-g_2 \sigma^2-g_3 \sigma^3\\
\left(\partial_t^2-\Delta+m^2_\omega\right)\omega_\mu&=&
~g_\omega j_\mu\\
\left(\partial_t^2-\Delta+m^2_\rho\right)\vec\rho_\mu&=&
~g_\rho \vec j_\mu\\
\left(\partial_t^2-\Delta\right)A_\mu&=&
~e j_\mu^{\rm em}.
\label{KGeq4}
\end{eqnarray}
In the mean-field approximation only the motion of the nucleons
is quantized, the meson degrees of freedom are described by 
classical fields which are defined by the nucleon densities and
currents. The single-particle spinors $\psi_i~(i=1,2,...,A)$
form the A-particle Slater determinant $|\Phi(t)\rangle$. 
The nucleons move independently in the classical meson fields,
i.e. residual two-body correlations are not included, and 
the many-nucleon wave function is a Slater determinant at 
all times. The sources of the fields in the Klein-Gordon
equations are calculated in the {\it no-sea}
approximation~\cite{VBR.95}:\ -~the scalar density
\begin{equation}
\rho_{\rm s}~=~\sum_{i=1}^A \bar\psi_i\psi_i,
\label{rho}
\end{equation}
-~the isoscalar baryon current
\begin{equation}
j^\mu~=~\sum_{i=1}^A \bar\psi_i\gamma^\mu\psi_i,
\label{current}
\end{equation}
-~the isovector baryon current
\begin{equation}
\vec j^{\,\mu}~=~\sum_{i=1}^A \bar\psi_i\gamma^\mu \vec \tau\psi_i,
\label{isocurrent}
\end{equation}
-~the electromagnetic current for the photon-field
\begin{equation}
j^\mu_{\rm em}~=~\sum_{i=1}^A
\bar\psi_i\gamma^\mu\frac{1-\tau_3}{2}\psi_i,
\label{emcurrent}
\end{equation}
where the summation is over all occupied states in the
Slater determinant $|\Phi(t)\rangle$. Negative-energy states
do not contribute to the densities in the {\it no-sea}
approximation for the stationary solutions, but their
contribution is implicitly included in
the time-dependent calculation ~\cite{VBR.95}.
The coupled system of equations
(\ref{dirac})--(\ref{KGeq4}) describes the time-evolution 
of A nucleons in the effective mean-field potential. The 
equations are equivalent to the equation of motion for the
one-body density operator $\hat \rho = \hat \rho(t)$ 
\begin{equation}
 i \hbar {\partial \over {\partial t}} \hat \rho = 
\left[ h_D, \hat \rho \right] ,
\label{eom}
\end{equation} 
with an initial condition for $\hat \rho$
\begin{equation}
\hat \rho(t_{in}) = \hat \rho_{in} .
\end{equation}
$h_D$ is the single-nucleon Dirac hamiltonian defined 
in Eq.~(\ref{dirac}). Starting from the self-consistent solution 
that describes the ground-state of the nuclear system, initial
conditions are defined to simulate excitations of giant resonances
in experiments with electromagnetic or hadron probes. 
For example, the one-body proton and neutron densities can 
be initially deformed and/or given some initial velocities.
The resulting dynamics can be described by the time-evolution
of the collective variables. In coordinate space for example,
these will be the multipole moments of the density distributions.
Of course, the dynamics of collective variables reflects 
the underlying single-nucleon motion in the self-consistent
potential. Since the Dirac hamiltonian depends on the nucleon
densities and currents through the solutions of the Klein-Gordon 
equations, it is obvious that the equations of motion are nonlinear.
For a specific choice of initial conditions the nuclear system could
enter into a chaotic regime of motion. The collective dynamics
that we describe is intrinsically classical, since it is 
formulated as a time-dependent initial value problem, 
rather than a boundary value problem. The 
single-nucleon wave functions on the other hand, satisfy the 
Pauli exclusion principle at all times, i.e. on the microscopic level
of single-nucleon motion the nucleus is a quantum system. The 
problem is therefore how to identify and quantify chaotic 
dynamics in an ensemble of nucleons that, described as quantum
objects on the microscopic level, display classical oscillations
of collective variables.  

\section{Time-series analysis}

Excitations of giant resonances result in damped 
harmonic/anharmonic density oscillations around the 
equilibrium ground-state of the nucleus. Since there are 
two types of nucleons in the nucleus, protons and neutrons,
a basic distinction is made between isoscalar and isovector 
oscillations. Isoscalar motion is characterized by proton 
and neutron densities oscillating in phase. The two densities
have opposite phases for isovector oscillations. In general, 
shape oscillations of the density can be represented as 
superposition of different multipoles: monopole, dipole,
quadrupole, etc. vibrations. Depending on the type and 
energy of the experimental probe, it is sometimes possible to 
selectively excite different multipoles. In addition to shape 
oscillations, spin degrees of freedom can be excited in 
giant resonances. The spin of the nucleon being  
naturally included in the relativistic framework, it
has been shown in Ref.~\cite{PVR.96} that the
time-dependent relativistic mean-field model provides
a consistent description of the spin-multipole resonances. 
Another type of excitations,
the Gamow-Teller resonances, include 
not only spin degrees of freedom, but also rotations in
isospin space, i.e. a neutron is transformed into a proton
or vice versa. In principle these resonances can also 
be described in the framework of relativistic mean-field
models, though the actual description of excitations that
involve so many degrees of freedom might become very complicated.
Therefore, in the present study we consider only the most 
simple situation: giant monopole resonances (GMR),but the results 
of our analysis should be valid also for shape oscillations of 
higher multipole order. These are just much more difficult 
to solve numerically. The equations of motion have to be 
integrated in two- or three-dimensional coordinate space,
and the numerical accuracy of our algorithms is simply 
not sufficient to obtain long time-series that are necessary
for an analysis of nonlinear dynamics. For excitations that
include spin and/or isospin degrees of freedom, the dynamics 
is more involved. Nevertheless one expects that the
present study will also provide some insight into the nonlinear
phenomena which occur in those more complicated excitations.     
    
The collective dynamical variables that characterize vibrations of a
nucleus are defined as expectation values of
single-particle operators in the time-dependent Slater
determinant $|\Phi(t)\rangle$ of occupied states: multipole moments 
that characterize the shape of the nucleus. 
In order to excite monopole oscillations, the spherical solution 
for the ground-state has to be initially compressed or radially
expanded by transforming the radial coordinate.
For isoscalar oscillations the monopole
deformations of the proton and neutron densities have the
same sign. To excite isovector oscillations, the initial
monopole deformation parameters of protons and neutrons
must have opposite signs.
For isoscalar monopole vibrations, the time-dependent monopole
moment is defined:
\begin{equation}
\langle r^2(t)\rangle ~=~\frac {1}{A}\langle \Phi(t) |r^2 |\Phi(t)\rangle,
\end{equation}
where A is the number of nucleons.
The corresponding isovector monopole moment is simply
\begin{displaymath}
< r^2_{\rm p}(t) > - < r^2_{\rm n} (t)>.
\end{displaymath}
Fourier transforms of the collective dynamical variable determine the
frequencies of eigenmodes.

In Fig.~\ref{figA} we display the time-series of monopole moments 
which represent the isoscalar and isovector 
oscillations in $^{208}$Pb. As in our calculation of 
Ref.~\cite{chaos.97}, the NL1 effective interaction has 
been used for the mean-field Lagrangian (effective masses
for the mesons and coupling constants of nucleons to
meson fields). This interaction reproduces the ground-state
properties of $^{208}$Pb, as well as experimental data 
on the energies of giant resonances: monopole, isovector 
dipole and isoscalar quadrupole. However, the precise values for 
the frequencies of the eigenmodes are not crucial in our study
of nonlinear dynamics. $^{208}$Pb is one of the most
studied nuclei, both experimentally and theoretically. 
The properties are well known, and we have selected $^{208}$Pb
because it is a heavy spherical system, with relatively 
little fragmentation of the modes, compared to lighter 
or deformed nuclei.    
The experimental isoscalar GMR energy in $^{208}$Pb is 
well established at $13.7\pm 0.3$ MeV, and  
the isovector GMR is at $26\pm 3$ MeV. The 
isoscalar mode displays regular undamped oscillations, 
while for the isovector mode we observe strongly 
damped anharmonic oscillations. Of course the time-series 
alone cannot determine whether the signal of the dynamical
variable displays characteristics of chaotic motion. 
In Fig.~\ref{figB} we show the corresponding Fourier power spectra
in logarithmic plots versus the excitation energy
$E = \hbar\omega$.  
There is very little spectral fragmentation 
in the isoscalar channel, and a
single mode dominates at the excitation energy of $\approx 11$ MeV.
The Fourier spectrum of the isovector mode is strongly fragmented. 
However, the main peaks are found in the energy region $25 - 30$ MeV, 
in agreement with the experimental data. The frequency of the 
isoscalar mode provides useful information about the underlying
dynamics, for example the compression modulus. On the other
hand, very little information can be inferred from the Fourier
spectrum of the isovector oscillations. This observation would
be consistent with the well known fact that nonlinear systems 
in the chaotic regime do not display any useful spectral content.
In order to extract more information from the time-series, 
methods of nonlinear analysis have to be used. 

For time-series that results from linear physical processes the 
Fourier analysis unfolds the characteristic frequencies which
are invariants of the dynamics, i.e. they classify the dynamics.
For nonlinear systems the corresponding analysis is somewhat 
more complicated. In order to reconstruct the dynamics of the
system from the time-series of a measured or calculated 
dynamical variable, one starts by reconstructing the phase space
using time delays. In this procedure there are two principal
quantities which have to be determined: the time delay 
and the dimension of the phase space on which the attractor 
unfolds. In this section we basically follow the prescriptions
of Ref.~\cite{Aba.93} for the reconstruction of the phase
space.

The time-series in Fig.~\ref{figA} have the form
\begin{equation}
x(n) = x(t_0 + n\tau_s)~~~~~~~~~~~n=0,1,2,...
\end{equation}
where $\tau_s$ is the sampling time. Since in our case the 
time-series is calculated by numerical integration of a set
of partial differential equations, the sampling time can be 
chosen arbitrarily. In order to define a coordinate system in 
which the structure of orbits in phase space can be described, 
time-lagged variables are used
\begin{equation}
x(n+T) = x(t_0 + (n+T) \tau_s),
\end{equation}
where T is some integer that defines the time delay. The 
collection of time-lagged variables defines a vector in 
$d$-dimensional phase space
\begin{equation}
\vec{y}(n) = \left[ x(n), x(n+T), x(n+2T), ... ,x(n+(d-1)T)\right].
\label{vector}
\end{equation}
In general, there is no unique prescription how to choose 
the optimal time lag T and the dimension of the space $d$.
The time delay should be chosen in such a way that
$x(n+jT)$ and $x(n+ (j+1)T)$ present two independent 
coordinates. If the time delay is too small, their numerical
values will be so close to make them practically indistinguishable;
if it is too large, the two coordinates will be completely 
independent of each other in a statistical sense, i.e. no dynamics
will connect their values. Of course, one also has to avoid that
the time delay coincides with a natural period of the system. 
The choice of the dimension of the reconstructed phase space $d_E$
(embedding dimension), depends on the dimension of the attractor.
$d_E$ must be sufficiently large so that the 
physical properties of the attractor are the same when computed
in time-lagged coordinates, and when computed in the physical
coordinates, which we do not know. For example, if two points
in the phase space are found close to each other, this should 
result from the underlying dynamics, and not from the small 
dimension of the phase space in which the dynamics is 
represented. The procedure is to embed the time-series in a 
$d_E$-dimensional phase space. The embedding dimension $d_E$ 
has to be equal or larger than the minimum number of dynamical
variables needed to model the system.

One possible way to choose the time delay is to consider the
linear autocorrelation function
\begin{equation}
C_L(T) = {{ \sum\limits_{n=1}^N 
\left[x(n+T) - \bar x\right]\left[ x(n) - \bar x\right]}\over
{\sum\limits_{n=1}^N \left[ x(n) - \bar x\right]^2} },
\end{equation}
where
\begin{equation}
\bar x = {1\over N} \sum\limits_{n=1}^N x(n) .
\end{equation}
By choosing the time lag to be the first zero of the 
autocorrelation function, $x(n+jT)$ and $x(n + (j+1)T)$ become,
on the average over the observation, linearly independent.   
The linear autocorrelation functions for isoscalar 
and isovector oscillations are shown in Fig.~\ref{figC}. 
The normalization is $C_L(T = 0) = 1$. In general, when
the time-series is irregular or chaotic, information about its past 
origins is lost. This means that $|C_L(T)| \to 0$ 
as $T \to \infty$, 
or the signal is only correlated with its recent past. 
For the isovector mode $|C_L(T)|$ indeed displays a 
much more rapid decrease, as compared to isoscalar
oscillations. The first zeros of the autocorrelation function
correspond to time delays of 27 fm/c and 13 fm/c, 
for the isoscalar and isovector
modes, respectively. These values could be used as time delays 
in the reconstruction of the corresponding phase spaces.
This method generally produces vectors in phase space with 
components that are, on the average, linearly independent.
For nonlinear systems a more appropriate method is to use
the average mutual information. This function can be 
considered as a generalization of the linear autocorrelation 
to nonlinear systems, and it tells us how much information
can be learned about a measurement at one time from a
measurement taken at another time. For a time series
x(n) and time-lag T, the average mutual information is 
defined 
\begin{equation}
I(T)= \sum\limits_{n=1}^N P(x(n),x(n+T))~{\rm log_2}~
\left[ {{ P(x(n),x(n+T))}\over {P(x(n)) P(x(n+T))}}\right] .
\end{equation}
The probability distribution $P(x(n))$ corresponds to 
the frequency with which any given value of $x(n)$ appears.
The joint distribution $P(x(n),x(n+T))$ corresponds to the 
frequency with which a unit box in the $x(n)$ versus 
$x(n+T)$ plane is occupied. The information functions calculated 
from the isoscalar and isovector time-series are displayed in
Fig.~\ref{figD}. We notice that on the average there is 
much more mutual information in the isoscalar signal. The
prescription is now to choose as time-lag the value for which
I(T) displays the first minimum~\cite{FS.86}. The curious 
result is that from the average mutual information we obtain
exactly the same time delays as from the linear autocorrelation
function: 27 fm/c for the isoscalar, and 13 fm/c for the 
isovector mode. 

In order to determine the embedding dimensions from the 
two time-series, we have used the method of false nearest
neighbors~\cite{Ken.92}. For each vector (\ref{vector}) in 
dimension $d$, we define a nearest neighbor $\vec {y}^{NN}(n)$.
The Euclidean distance between the two vectors is
\begin{equation}
R_d^2(n) = [x(n) - x^{NN}(n)]^2 + ... + 
[x(n+(d-1)T) - x^{NN}(n+(d-1)T)]^2
\end{equation}
Of course the two vectors are nearest neighbors if this distance
is small, in a sense that we will define shortly. In 
dimension $(d+1)$ the distance between the two vectors becomes
\begin{equation}
R_{d+1}^2(n) = R_d^2(n) + [x(n+dT) - x^{NN}(n+dT)]^2.
\end{equation}
If the two points were nearest neighbors in dimension $d$, but 
now we find that $R_{d+1}^2(n)$ is large compared to $R_d^2(n)$,
this must be due to the projection from some higher dimensional
attractor down to dimension $d$. By going from dimension $d$ to
$(d+1)$, we have shown that the two points were "false nearest
neighbors". If the ratio 
\begin{equation}
{{| x(n+dT) - x^{NN}(n+dT)|} \over {R_d(n)}}
\end{equation}
is larger than some threshold, the assertion that the
two vectors are nearest neighbors is false. The fact that they
are found to lie close to each other in dimension
$d$ is not a property of the dynamics of the system, but the
result of projecting the dynamics onto a phase space of too low
dimension. The embedding dimension is now determined with
a simple procedure. First we decide what value for the 
Euclidean distance should be taken as small. This generally
depends on the data set that we are analyzing. For 
all vectors in the phase space of dimension $d$ we count the 
number of nearest neighbors. Then we have to determine the 
percentage of nearest neighbors that turn out to be false when
going to dimension $(d+1)$. The minimal necessary embedding 
dimension $d_E$ is selected to be the one for which the 
percentage of false nearest neighbors goes to zero. For the 
isoscalar and isovector time-series, the false nearest 
neighbors are displayed in Fig.~\ref{figE} as functions 
of the phase space dimension. The percentage of false nearest
neighbors goes to zero for $d_E=3$ (isoscalar mode), and
for $d_E=4$ (isovector mode). These values are taken as embedding
dimensions for the reconstruction of the corresponding phase spaces.

The reconstructed phase space can be represented by the 
recurrence plot. By embedding the time-series we create 
a sequence of vectors 
\begin{displaymath}
\vec {y}(n) = \left[ x(n), ... ,x(n+(d_E - 1)T) \right]
\end{displaymath}
in the phase space of dimension $d_E$. In our example 
the time delay is 27 fm/c for the isoscalar, and 13 fm/c 
for the isovector mode. The corresponding embedding dimensions 
are $d_E=3$ and $d_E=4$, respectively. We can calculate the 
distance between any two points in the phase space
\begin{displaymath}
\delta(m,n) = |\vec{y}(m) - \vec{y}(n)| . 
\end{displaymath}
To construct the recurrence plot we choose some distance $r$, and
ask when $|\vec{y}(m) - \vec{y}(n)| < r$. $m$ is placed on the 
horizontal axis, $n$ on the vertical axis, and a dot is placed
at the coordinate $(m,n)$ if $|\vec{y}(m) - \vec{y}(n)| < r$.
For a periodic signal the recurrence plot displays a series
of stripes at 45 degrees. If a time-series is chaotic, the 
recurrence plot has a more complicated structure. It is non-uniform
and boxes of dense points appear along the diagonal. Of course, in 
all recurrence plots there is a stripe on the diagonal $m=n$. 
The recurrence plots for the phase spaces of the isoscalar and
isovector time-series are shown in Fig.~\ref{figF}. We notice
a pronounced difference between the two modes. For the 
isoscalar mode the recurrence plot displays a pattern 
representative for regular oscillations, with stripes separated 
by a distance that corresponds to the period of oscillations.
On the other hand, the recurrence plot for the isovector mode
indicates non-stationarity.

The number of dots in a recurrence plot tells how many times
the phase space trajectory came within distance $r$ of a 
previous value. A measure of the density of dots is provided
by the correlation integral. If the dynamics of a system is 
deterministic, the ensemble of phase space trajectories 
converges towards an invariant subset of the phase space - 
the attractor. For chaotic dynamics the attractor has fractional
dimension, whereas the dimension is integer for regular dynamics. The
correlation dimension of the attractor can be numerically evaluated from 
the correlation integral. If there are N points $\vec{y}(n)$ in the
reconstructed phase space of dimension $d$, we can compute all 
distances $|\vec{y}(m) - \vec{y}(n)|$. The correlation integral
is defined~\cite{GP.83}
\begin{equation}
C_2 (r) = {2\over {N(N-1)}} \sum\limits_{m\neq n}^N 
\Theta(r - |\vec{y}(m) - \vec{y}(n)|) ,
\end{equation}
for a distance $r$ in phase space. $\Theta(x) = 0$ if  
$x < 0$ and $\Theta(x) = 1$ for $x > 0$. In a certain range of 
$r$, the scaling region, $C_2(r)$ behaves like
\begin{equation}
C_2(r) = r^d.
\end{equation}
The correlation dimension $D_2$ is determined by the slope of the 
$log~C_2(r)$ versus $log~r$. It is defined as the slope of the 
plot in the $r \to 0$ limit. The dimension of the attractor
can be determined by plotting $log~C_2(r)$ versus $log~r$ 
for a set of increasing dimensions of the phase space. 
As the embedding dimension increases, the correlation 
dimension $D_2$ should saturate at a value equal to the 
attractor's correlation dimension. The logarithm of the correlation
integral is plotted in Fig.~\ref{figG} for the isoscalar  
and isovector modes, for a set of increasing dimensions 
(direction of the arrow). The corresponding correlation
dimensions are displayed in Fig.~\ref{figH}. We notice that
for the isoscalar mode, for $d\geq 3$, the correlation 
dimension saturates at $D_2 = 2$. The integer value for the 
dimension of the attractor indicates
regular dynamics. For the isovector mode the correlation 
dimension does not saturate, but slowly increases to some 
fractional value between 2 and 3. The fractional dimension
of the attractor would imply chaotic or stochastic dynamics.

\section{Information-theoretic functionals}

In this section we continue with the analysis of nonlinear
dynamics in time-series of giant monopole oscillations.
The identification and quantification of the underlying 
regular or chaotic dynamics will be based on the evaluation of 
information-theoretic functionals. We start with the information 
entropy of a physical system. For classical systems the entropy
is defined on the phase space density, and represent the missing
information about which fine-grained cell of the phase space
the system occupies. For a quantum system the 
information entropy can be defined on the density operator. 
The von Neumann entropy of the one-body density operator 
$\hat D$ (measured in bits)
\begin{equation}
S = - tr(\hat D~{\rm log_2}~\hat D) = - \sum\limits_j 
\lambda_j~{\rm log_2}~\lambda_j ,
\end{equation} 
where $\lambda_j$ are eigenvalues of the density operator
(occupation probabilities), 
and the entropy can be interpreted as the missing information 
about which eigenvector the system occupies.
Entropy is conserved under Hamiltonian dynamical evolution, 
both classically and for a quantum system. Classically this 
is a consequence of the conservation of phase space volume, in
quantum mechanics it follows from the unitarity of Hamiltonian
evolution. However, if the system is coupled to a perturbing 
environment, the interaction generally changes the system's 
entropy. In our model only the motion of nucleons, protons
and neutrons, is quantized. The equation of motion (\ref{eom})
describes the time evolution of the one-body density in the 
time-dependent meson fields. The mean-fields play the 
role of the environment, the self-consistent interaction 
of the nucleon with the meson fields determines the nonlinear
dynamics. This implies that, if from the nucleon densities 
we define some time-dependent entropy functionals, their
time evolution might contain useful information
about the dynamics of the system. For example, we can define
the information entropy functional
\begin{equation}
S(t) = - \int \rho(\vec r,t)~{\rm log_2}~\rho(\vec r,t)~d^3 r ,
\label{entropy}
\end{equation}
where $\rho$ is the vector density (0-th component in 
Eq.~\ref{current}). Since we consider both 
isoscalar and isovector motion, the density in (\ref{entropy})
can be the neutron or the proton density, or the total
nucleon density.

In Fig.~\ref{figI} we display the time-dependent total (a),
neutron (b), and  proton (c) entropies (\ref{entropy}) for the isoscalar 
and isovector monopole oscillations in $^{208}$Pb. In addition, 
in all three cases we include the entropy that results from the 
time evolution of the system that has not been excited in any 
way. This ground state entropies provide a measure of the numerical
accuracy of the integration algorithm. We notice that for the 
isoscalar mode the entropies display regular oscillations 
which reflect the exchange of energy between the nucleons and 
the meson fields. The oscillations are identical to those of the
dynamical variable, the isoscalar monopole moment in Fig.~\ref{figA}.
For the isovector mode the entropies, in addition to somewhat
more complicated oscillations, slowly decrease to the values 
that are characteristic for the ground-state of the nucleus. 
This decrease in entropy is caused by the strong mean-field 
damping of the isovector mode, i.e from the collisions of the 
nucleons with the moving wall of the nuclear potential generated
by the self-consistent meson fields. In the isovector mode the
protons and neutrons effectively move in two self-consistent
potentials that oscillate out of phase, and which in this way inhibit 
the resonance. To extract the information content of the 
time-dependent entropies (\ref{entropy}), we have calculated the Fourier 
transforms. For the total, neutron and proton entropies, the 
Fourier power spectra are displayed in Figs.~\ref{figJ} and 
\ref{figK}, for isoscalar and isovector oscillations, 
respectively. For the isoscalar mode the information content
of the entropy is exactly the same as that of the dynamical 
variable, the monopole moment: a single mode dominates, at 
a frequency which corresponds to the excitation energy of 
the giant monopole resonance. This is not surprising, if one
considers that the monopole moment is defined with an integral
identical to the one that defines the entropy in (\ref{entropy}), 
except that $-{\rm log_2}~\rho$ is replaced by $r^2$. The situation 
is different for the isovector mode (Fig.~\ref{figK}). The  
entropy contains more information than the dynamical variable.
In addition to the frequencies in the region of isovector 
monopole resonances ($25 - 30$ MeV), there are strong 
peaks at the frequency of the isoscalar resonance. They
are related to the compressibility modulus
of the nuclear matter. The entropy of the total density
therefore contains information about both modes, 
but now we notice that the Fourier spectra for the neutron 
and proton entropies are different. For neutrons the peaks 
in the region of isovector excitations are strongly suppressed,
and there is fragmentation at the frequency of the isoscalar mode.

The radius of a heavy spherical nucleus like $^{208}$Pb is 
$\approx 5-6$ fm. The giant multipole resonances represent
collective oscillations of the proton and neutron densities,
and therefore
provide excellent physical examples for the analysis of 
systems that have spatial as well as temporal structure.
For a nonlinear system in chaotic regime, we might consider 
the influence of spatial motion on temporal chaos. We  
ask what are the spatial correlations in a finite system that 
displays chaotic oscillations of a collective dynamical variable. 
Consider, for example, the conditional entropy defined from a 
two-body total density
\begin{equation}
S_2(t) = - \int \rho^2(\vec r,\vec r~',t) 
 ~{\rm log_2}~\left[ {{\rho^2(\vec r,\vec r~',t)}\over
{\rho(\vec r,t) \rho(\vec r~',t)}}\right]
 ~d^3 r~d^3 r' ,
\label{entropy2}
\end{equation}
where the two-body density matrix is defined from the 
Slater determinant of occupied states
\begin{eqnarray}
\rho^2=\sum_{kijl}~\langle i|\hat{\rho}(\vec r) |j \rangle  
\langle k| \hat{\rho}(\vec r~') |l \rangle
\langle\Phi(t)|a_i^+ a_k^+ a_l a_j|\Phi(t)\rangle.
\label{doubdens2}
\end{eqnarray}
In coordinate representation the expression becomes
\begin{eqnarray}
&&\langle\Phi(t)|:\hat{\rho}(\vec r) 
\hat{\rho}(\vec r~'):|\Phi(t)\rangle =
\rho(\vec r) \rho(\vec r~') -\nonumber\\
&&\sum^{Z}_{ij}\psi^{+}_{i}(\vec r) \psi_{j}(\vec r)
\psi^{+}_{j}(\vec r~') \psi_{i}(\vec r~') -
\sum^{N}_{ij}\psi^{+}_{i}(\vec r) \psi_{j}(\vec r)
\psi^{+}_{j}(\vec r~') \psi_{i}(\vec r~').
\nonumber\\
\label{doubdens6}
\end{eqnarray}    
$\psi_{i}(\vec r)$ denotes the single nucleon Dirac spinor,
and Z (N) is the number of protons (neutrons).
The conditional entropy (\ref{entropy2}) should provide a 
measure of two-body spatial correlations. For some kind of 
collective motion, regular or chaotic, this function contains the
following information: how much are the oscillations of nucleon
density at some point in space determined by the oscillations 
at some other point in the system, i.e. what are the correlations 
between oscillations of nucleon density at various points in
the finite system.

The time-dependent entropies (\ref{entropy2}) that corresponds 
to isoscalar and isovector oscillations are shown in 
Fig.~\ref{figL}. They are compared with the value 
that results from the time-evolution of the system that 
has not been excited (time-dependent entropy of the ground state).
For the isoscalar mode, regular modulated oscillations are 
observed. Comparing also with the reference ground-state entropy,
we notice how the numerical accuracy affects the results for 
long times of integration (T $> 2000$ fm/c). The 
entropy that corresponds to the isovector mode is much lower and more
irregular at the beginning, but it eventually approaches values 
comparable to those of the isoscalar mode. Similar to
the entropy defined on the one-body density operator,
this behavior reflects the strong mean-field damping of the
isovector oscillations. The information contents of the "two-body"
entropies are shown in the corresponding Fourier power spectra in 
Fig.~\ref{figM}. For the isoscalar mode we again find that the
entropy contains the same information as the dynamical variable,
a single mode at the frequency of the giant resonance. This means
that there is a high degree of two-body correlations for the 
isoscalar mode, the nucleon density oscillates with the same
frequency at all points in the nuclear system. For the 
isovector mode we do not find any useful information in the
Fourier spectrum. There is a highly fragmented structure in the
region of the isoscalar giant resonance, but in addition we 
find strong peaks in the very low frequency region. This 
result indicates that there is very little spatial correlation
for the isovector oscillations of the nucleon density, or
that the nonlinear nuclear system oscillates in a regime 
for which the Fourier spectrum of the conditional entropy 
(\ref{entropy2}) does not contain useful information. 

In the previous section, we have used the average mutual information
function to determine the time delay for the reconstruction of 
the phase space. 
This function quantifies the information that is contained in 
the signal, at some moment in time, about the value of the 
dynamical variable at other times. Since we describe isoscalar
and isovector oscillations, i.e. we distinguish between 
proton and neutron components of the system, we might ask
how much information is contained in the dynamical variable
of the neutron distribution, about the proton subsystem, 
and vice versa. The two dynamical variables in this example
are the mean square radii of the two distributions. We
define the information function
\begin{equation}
I_{\pi(\nu)}(\epsilon) = - \sum\limits_i P_i (x_{\pi(\nu)})~
{\rm log_2}~P_i (x_{\pi(\nu)}) .
\end{equation}
The signal $x$ is quantized in units of $\epsilon$.
The probability distribution $P_i(x)$ corresponds to 
the frequency with which any given value of x appears
in the box $i$ of dimension $\epsilon$. The sum is over occupied 
boxes of dimension $\epsilon$, in the one-dimensional embedding of 
the time-series. For two time-series, the corresponding joint
information function is
\begin{equation}
I_{\pi,\nu} (\epsilon) = - \sum\limits_{i,j} P_{i,j}(x_\pi,y_\nu)
 ~{\rm log_2}~ P_{i,j}(x_\pi,y_\nu)
\end{equation}
The joint distribution $P_{i,j}$ corresponds to the 
frequency with which a box $(i,j)$ (linear dimension $\epsilon$)  
in the $x_\pi$  versus $y_\nu$ plane is occupied. 
The average amount of information about the variable $y$ 
that the variable $x$ contains is quantified by the 
mutual information~\cite{Pal.93}
\begin{equation}
M_{x,y} (\epsilon) = I_x (\epsilon) + I_y (\epsilon)
 - I_{x,y} (\epsilon)
\label{info}
\end{equation}
Clearly, the mutual information vanishes if 
$ P_{i,j}(x,y) = P_i(x) P_j(y)$, i.e. if $x$ and $y$ are 
statistically independent. The precise value of the mutual 
information will of course depend on the size of the box
$\epsilon$, but one should try to find a region of values
for $\epsilon$ in which $ M_{x,y} (\epsilon)$ does not vary 
appreciably. 

In our example of giant monopole resonances, the variable
$x$ corresponds to the mean square radius of the proton
distribution, and $y$ to that of the neutron distribution.
The mutual information functions (in units of bits) are 
displayed in Fig.~\ref{figN}, for the isoscalar and 
isovector oscillations. The acceptable values for 
$\epsilon$ depend on the sampling of the time-series. 
For $\epsilon\leq 0.2$ fm$^2$ the probability distribution 
functions cannot be properly determined, there are many 
empty boxes, and the calculated mutual information is 
not useful. For larger values of $\epsilon$ the calculated mutual 
information changes very slightly, practically  
with the same slope for isoscalar and isovector modes.
Of course, the principal result is the comparison
between the two modes: the average amount of information that
$\langle r^2\rangle$ of the proton density contains about
the dynamical variable of the neutron distribution is more 
than a factor three larger for the isoscalar mode.

Another interesting possibility is to consider the 
mutual information as function of the spatial coordinate. 
Instead of using as dynamical variables integrated quantities
like the mean square radii, we can follow the time evolution 
of the proton and neutron densities at various points 
along the radial axis (the motion is spherically symmetric).
The dynamical variables $x$ and $y$ will be the values of
the proton and neutron densities at each point in space, 
and we can plot the average mutual information of the 
densities as function of the radial coordinate. The results are
shown in Fig.~\ref{figO}. In addition to $^{208}$Pb, we
also display the mutual information for $^{16}$O and $^{40}$Ca.
These two spherical nuclei are smaller, but have the advantage
to contain identical numbers of protons and neutrons. In all
three nuclei the mutual information of the proton and
neutron density is much higher for the isoscalar mode. In fact,
for $^{16}$O and $^{40}$Ca, the mutual information for the 
isovector mode practically vanishes, and no radial dependence
is observed. It is somewhat higher for $^{208}$Pb, and with 
some modulation, slowly decreases from the center of the nucleus
towards the surface. The isoscalar mode displays a very interesting 
radial behavior of the mutual information. It is high in the
nuclear volume, but there is also a pronounced minimum at the 
radius that corresponds to the surface of the nucleus. This means
that there is little correlation between proton and neutron 
densities in the surface region, they oscillate almost
independently. The mutual information increases again beyond the
ground-state radius of the nucleus, but in this asymptotic 
region the densities rapidly decrease to zero. Of course, the 
behavior in the surface region is not completely unexpected.
The nucleons at the surface are less bound, and the effective
compression modulus of the surface region is different from that
in the volume of the nucleus. For example, the nucleus 
$^{208}$Pb contains 82 protons and 126 neutrons. However, 
due to the combined effects of Coulomb repulsion between protons,
and the Pauli exclusion principle, the protons occupy practically
the same volume as the neutrons. Yet the dynamics for the two
types of nucleons seem to be very different in the surface
region. The slowly vibrating self-consistent potentials in which
the protons and neutrons move, do not average on the surface in the
same way as in the bulk region. It is very interesting
how the details of the underlying nonlinear dynamics emerge
in the radial behavior of the mutual information function.  
\section{Conclusions}

In the present work we have used the time-dependent relativistic
mean-field model to analyze the nonlinear dynamics of giant 
resonances in atomic nuclei. The characteristic properties
of these collective excitations vary smoothly with the size of
the nucleus, and therefore a self-consistent mean-field approach 
provides a consistent description of nucleon dynamics. In particular, 
we have analyzed the time-series of dynamical variables that
characterize the giant monopole resonances: isoscalar (proton and
neutron densities oscillate in phase), and isovector (proton 
density oscillates against the neutron density). The nucleons
move in the effective self-consistent single nucleon potentials,
and the equations of motion describe the time evolution of the 
one-body density. Since the time-dependent potentials are calculated
in a self-consistent way, the model of the nuclear system is intrinsically
nonlinear, and chaotic motion is expected for specific 
initial conditions.

From the time-series of isoscalar and isovector monopole moments of
$^{208}$Pb, we have reconstructed the corresponding phase spaces. The  
time delays have been calculated from the average mutual information,
and the embedding dimensions determined by the method of false
nearest neighbors. The reconstructed phase spaces have been 
represented by recurrence plots. We have found that for the 
isoscalar mode the recurrence plot displays a pattern characteristic
for regular oscillations, while for the isovector mode it indicates
non-stationarity. From the reconstructed phase-spaces we have also
calculated the correlation integrals and the corresponding 
correlation dimensions. As a function of the embedding dimension of 
the phase space, the correlation dimension $D_2$ saturates at the 
integer value 2 for the isoscalar mode. On the other hand, a 
fractional correlation dimension is found for the isovector 
oscillations. The results confirm our conclusions from 
Ref.~\cite{chaos.97} that the motion of the collective coordinate 
is regular for isoscalar oscillations, and that it becomes
chaotic when initial conditions correspond to the isovector mode.

The nonlinear dynamics of giant resonances has also been analyzed
in the framework of information-theoretic functionals. For the 
time-dependent one-body nucleon densities, we have calculated 
the von Neumann information entropy functionals. 
The Fourier analysis has shown that the  
entropy of the isoscalar mode contains the same information as
the dynamical variable. The structure is more complicated 
for the isovector mode, for which peaks are found both in the
regions of isoscalar and isovector eigenfrequencies. The 
spatial correlations have been described with a time-dependent
conditional entropy defined from a two-body nucleon density. 
This function enables the study of the influence of spatial 
motion on temporal chaos. From the dynamical variables that
characterize the proton and neutron distributions, i.e. the
mean square radii, we have calculated the average mutual 
information for the isoscalar and isovector modes. The average
information that is contained in the collective dynamical variable 
of the proton density, about the neutron density, is more than
a factor three larger for the isoscalar mode. We have also 
analyzed the mutual information between proton and neutron
densities as a function of the spatial coordinate. It has been 
shown that, not only is the average mutual information much 
higher for the isoscalar mode, but it also displays an 
interesting radial dependence which reflects the differences
in the dynamics of the monopole motion in the volume and 
on the surface of the nucleus.

The results of the present analysis, as well as those of 
Ref.~\cite{chaos.97}, have shown that giant resonances in 
nuclei provide excellent examples for the study of regular
and chaotic dynamics in quantum systems. In addition, the 
finite spatial extension of nuclei enables the analysis
of spatio-temporal behavior in nonlinear dynamical systems.
And yet we have only examined the most simple modes of collective
motion: monopole oscillations. More complicated excitations,
especially those involving spin and isospin degrees of freedom,
would certainly disclose more interesting properties of the 
underlying nonlinear dynamics. 
       
\newpage
{\bf Acknowledgments}

This work has been supported in part by the
Bundesministerium f\"ur Bildung und Forschung under
contract 06~TM~875. 


\newpage

\centerline{\bf Figure Captions}
\bigskip

\begin{figure}
\caption{Time-dependent isoscalar $<r^2>$ (a), and
isovector $<r_{\rm p}^2> - <r_{\rm n}^2>$ (b) monopole moments
for $^{208}$Pb.}
\label{figA}
\end{figure}

\begin{figure}
\caption{Fourier power spectra for the isoscalar (a), and
isovector (b) monopole oscillations in $^{208}$Pb.}
\label{figB}
\end{figure}

\begin{figure}
\caption{Linear autocorrelation functions
for isoscalar (a), and isovector (b) oscillations.}
\label{figC}
\end{figure}

\begin{figure}
\caption{Average mutual information as a function of
time delay for monopole oscillations.}
\label{figD}
\end{figure}

\begin{figure}
\caption{Percentage of false nearest neighbors as a 
function of phase space dimension, 
for isoscalar (a), and isovector (b) modes.} 
\label{figE}
\end{figure}

\begin{figure}
\caption{Recurrence plots for the time-series of 
isoscalar (a), and isovector (b) monopole oscillations.} 
\label{figF}
\end{figure}

\begin{figure}
\caption{Sequence of correlation integrals for 
 isoscalar (a), and isovector (b) monopole oscillations.
log $C_2(r)$ $vs$ log $r$ plots are displayed for a 
sequence of embedding dimensions $d = 1,...,8$.} 
\label{figG}
\end{figure}

\begin{figure}
\caption{Correlation dimension $D_2$ as function of 
embedding dimension, for
isoscalar and isovector oscillations.} 
\label{figH}
\end{figure}

\begin{figure}
\caption{Time-dependent entropy functionals (\ref{entropy}) for 
isoscalar and isovector monopole motion. We display the 
total entropy of the nucleon system (a), the neutron (b),
and the proton entropy (c). Solid curves correspond to 
isoscalar oscillations, dashed curves to isovector 
oscillations. The thick solid curves are the reference 
ground-state entropies.}
\label{figI}
\end{figure}

\begin{figure}
\caption{Fourier power spectra of the total (a), neutron (b)
and proton (c) entropies, for isoscalar monopole oscillations.}
\label{figJ}
\end{figure}

\begin{figure}
\caption{Fourier power spectra of the total (a), neutron (b)
and proton (c) entropies, for isovector monopole oscillations.}
\label{figK}
\end{figure}

\begin{figure}
\caption{Time-dependent conditional entropies (\ref{entropy2})  
for isoscalar and isovector monopole motion. The thick solid
line is the reference ground-state entropy.}
\label{figL}
\end{figure}

\begin{figure}
\caption{Fourier power spectra of the conditional
entropies (\ref{entropy2}) for isovector (a), and 
isoscalar (b) monopole oscillations.}
\label{figM}
\end{figure}

\begin{figure}
\caption{Mutual information (\ref{info}) between the time-dependent
mean square radii of the proton and neutron density distributions.
The two curves that correspond to isoscalar and isovector 
oscillations, are plotted as functions of the size of the box
in the linear embedding of the time-series.}
\label{figN}
\end{figure}

\begin{figure}
\caption{Radial dependence of the mutual information between
proton and neutron density distributions. Results for 
$^{16}$O, $^{40}$Ca and $^{208}$Pb are displayed. Solid 
curves correspond to isoscalar oscillations, dashed curves
to isovector oscillations.}
\label{figO}
\end{figure}

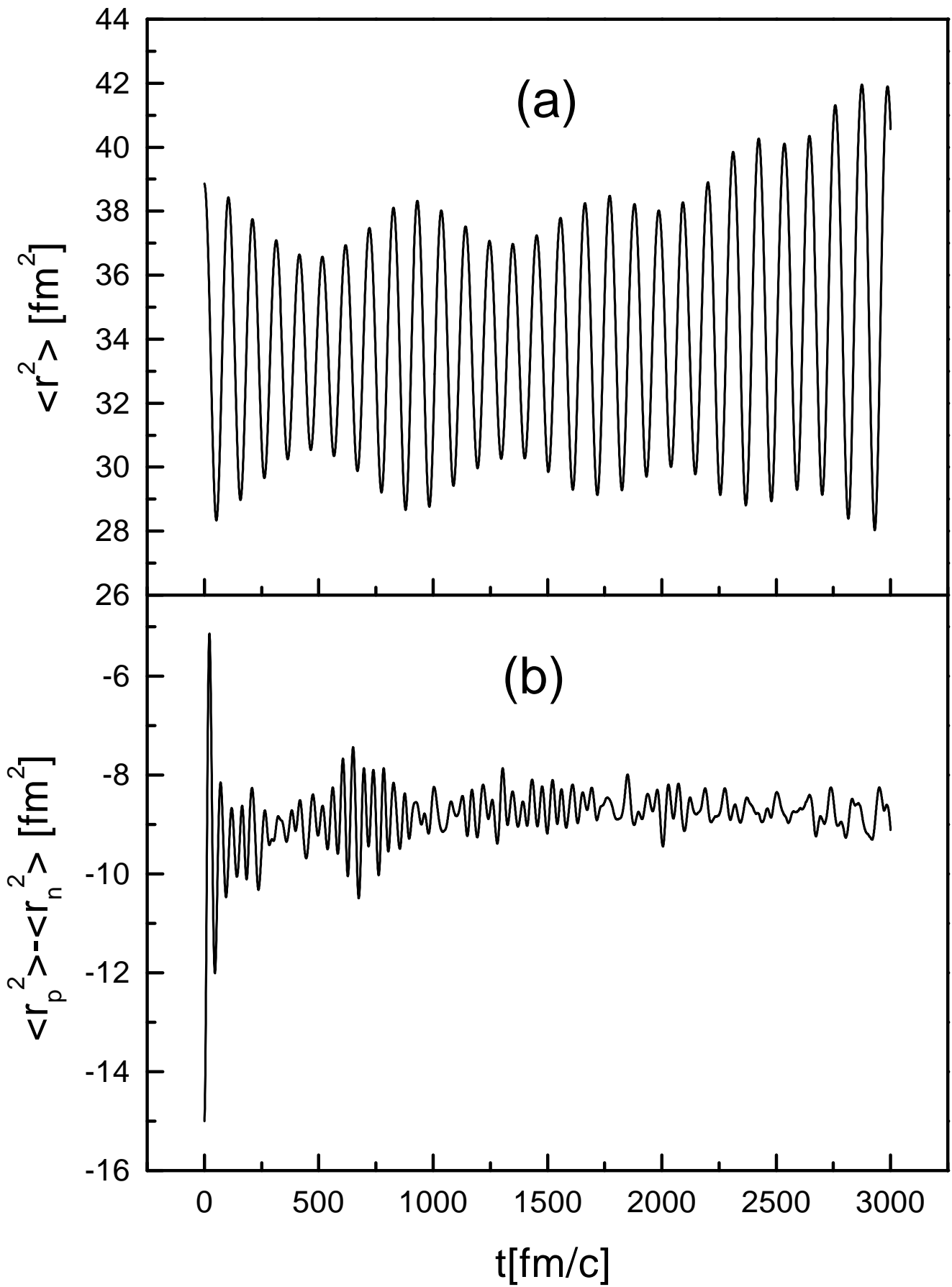
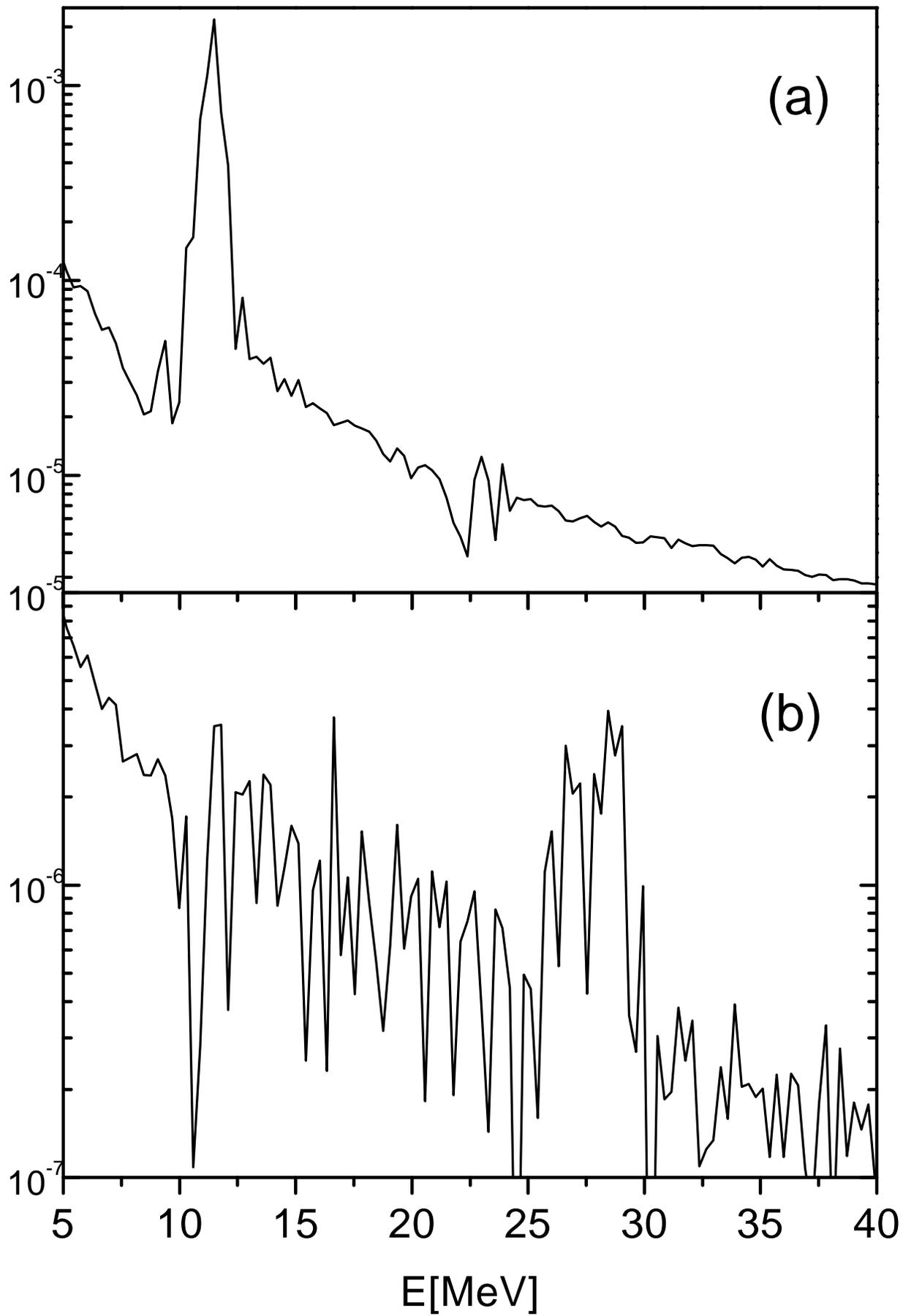
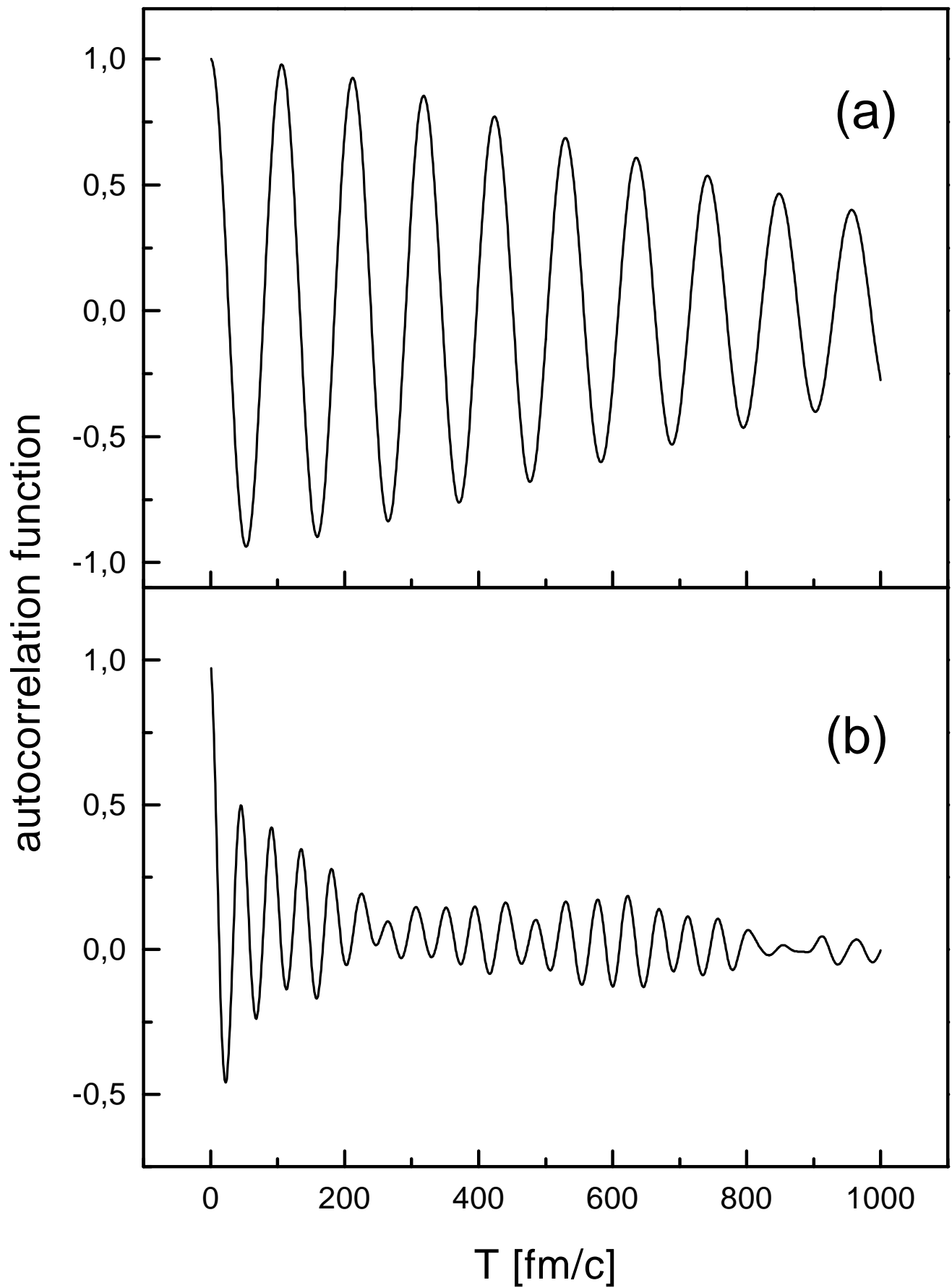
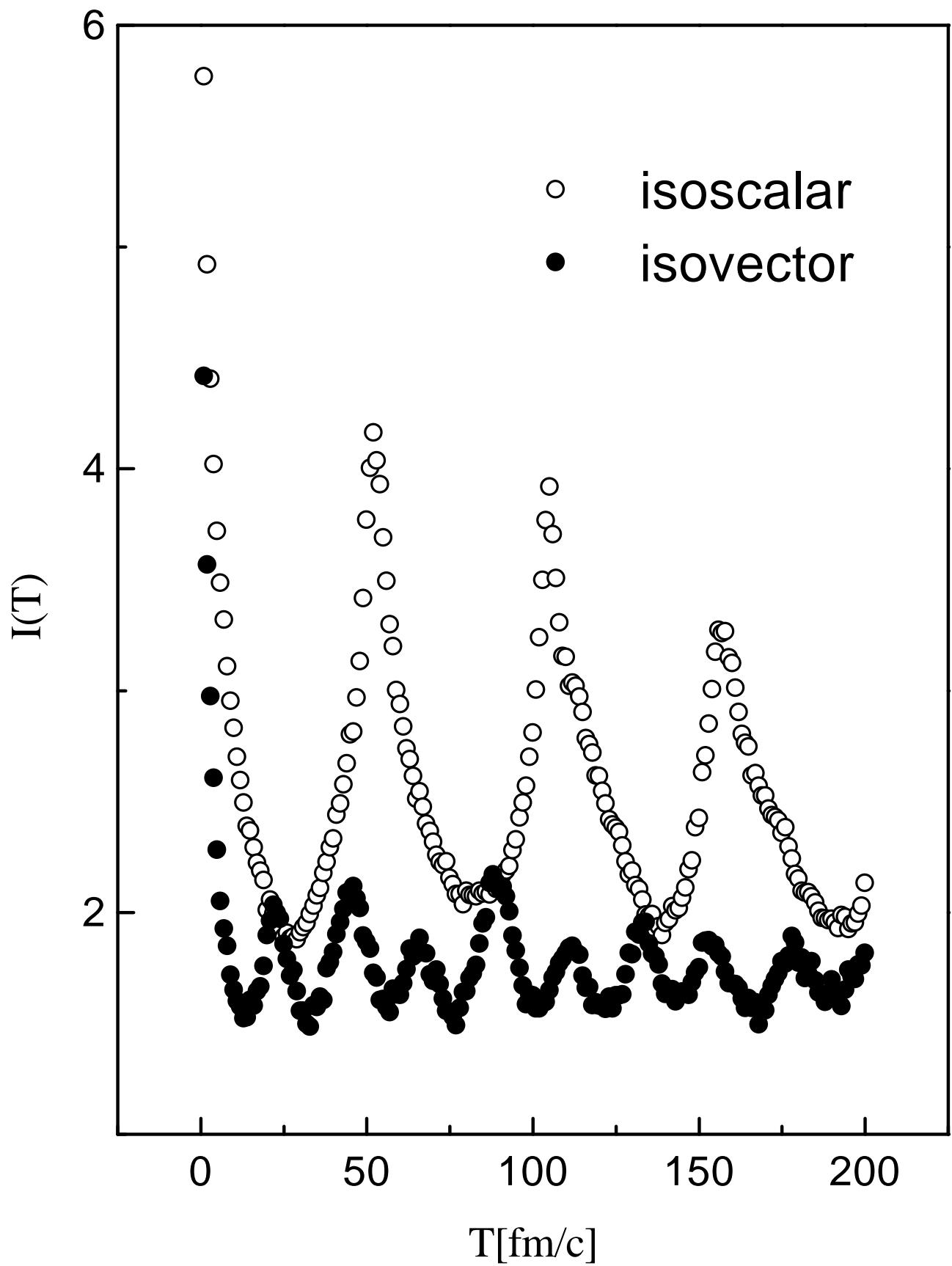
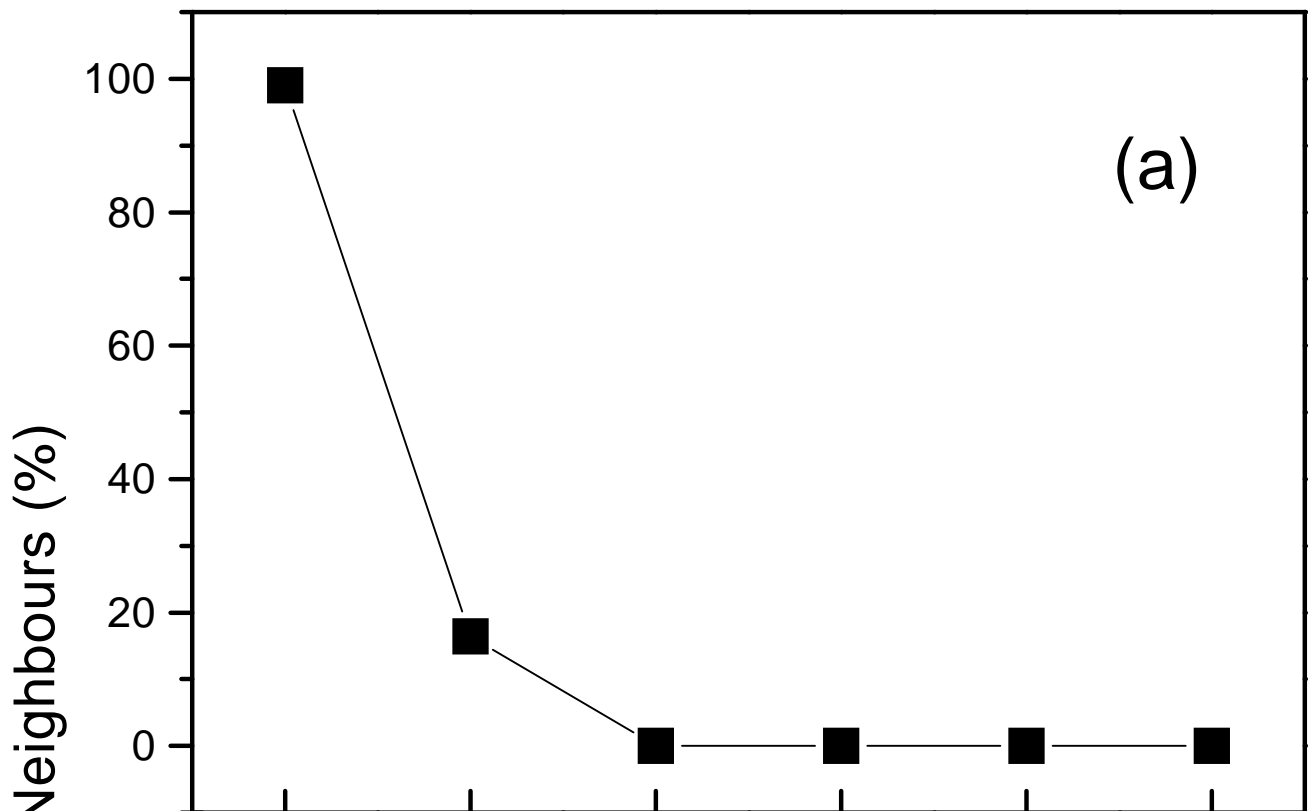
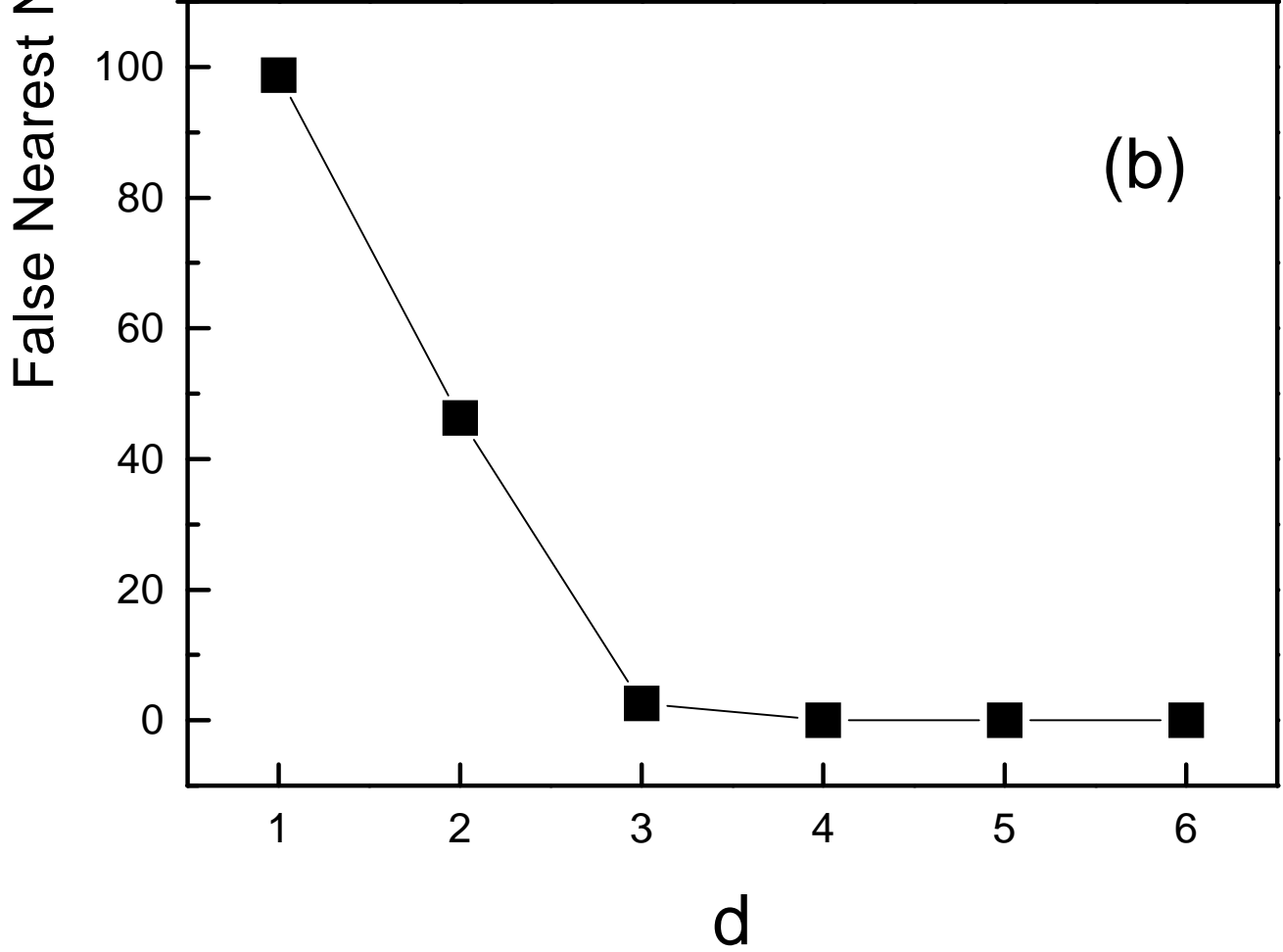
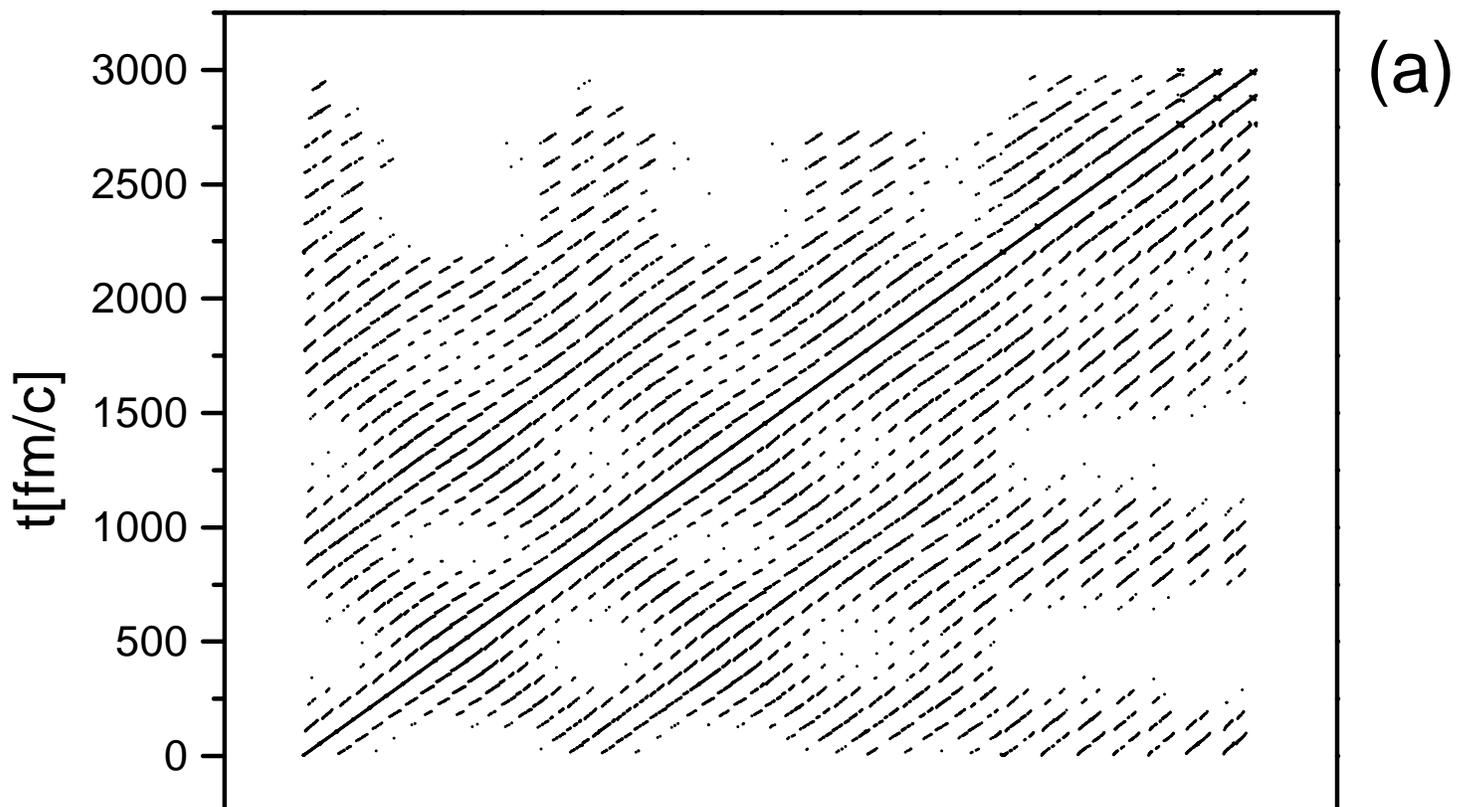
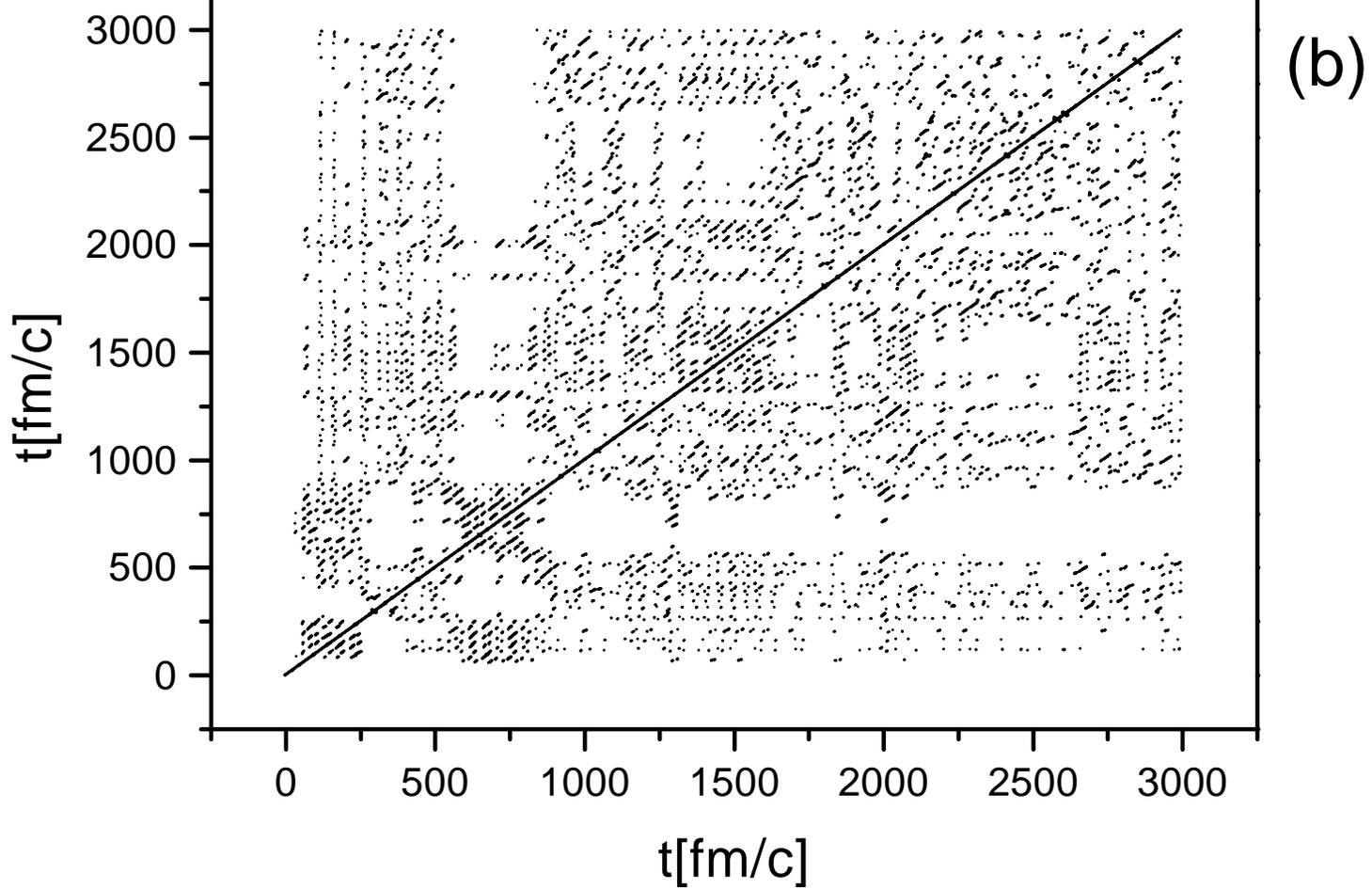
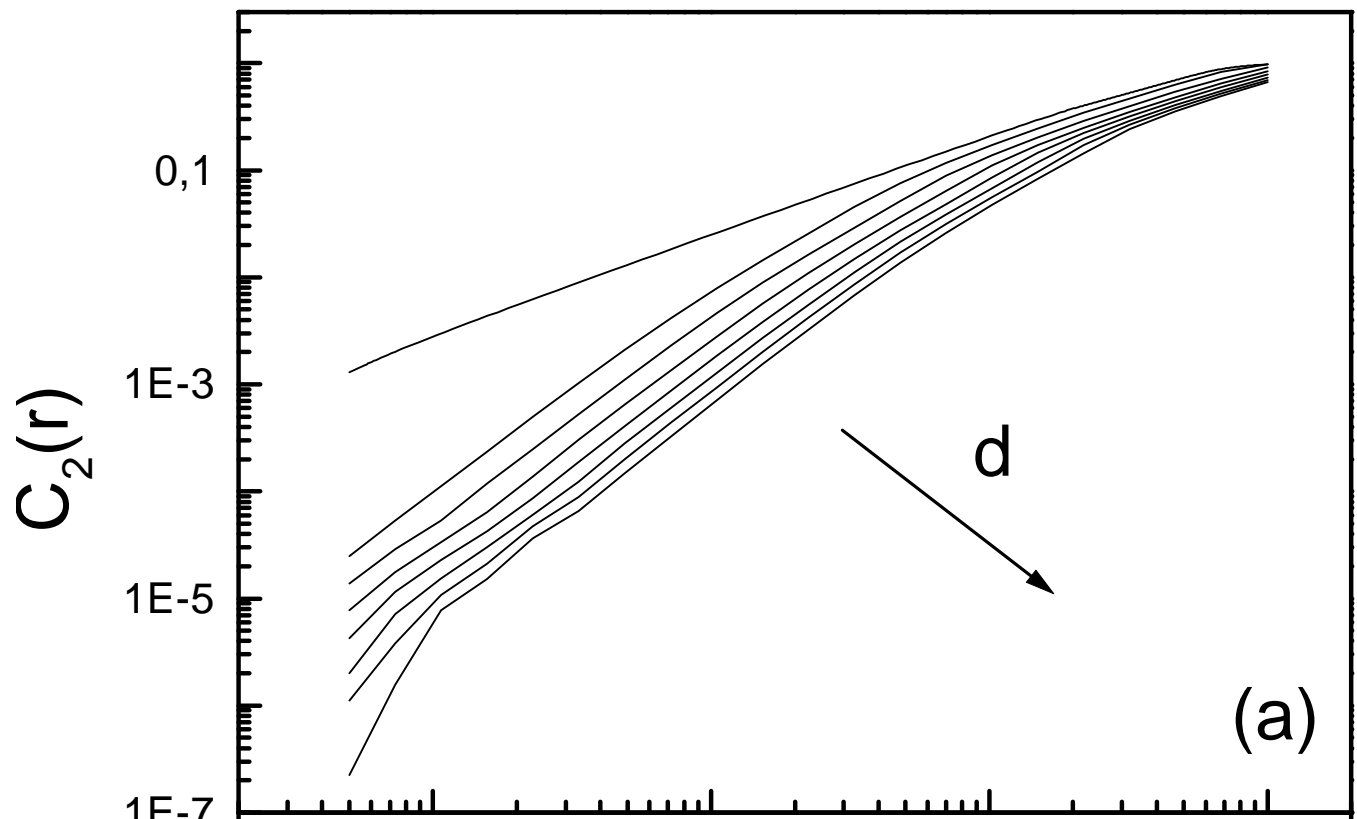
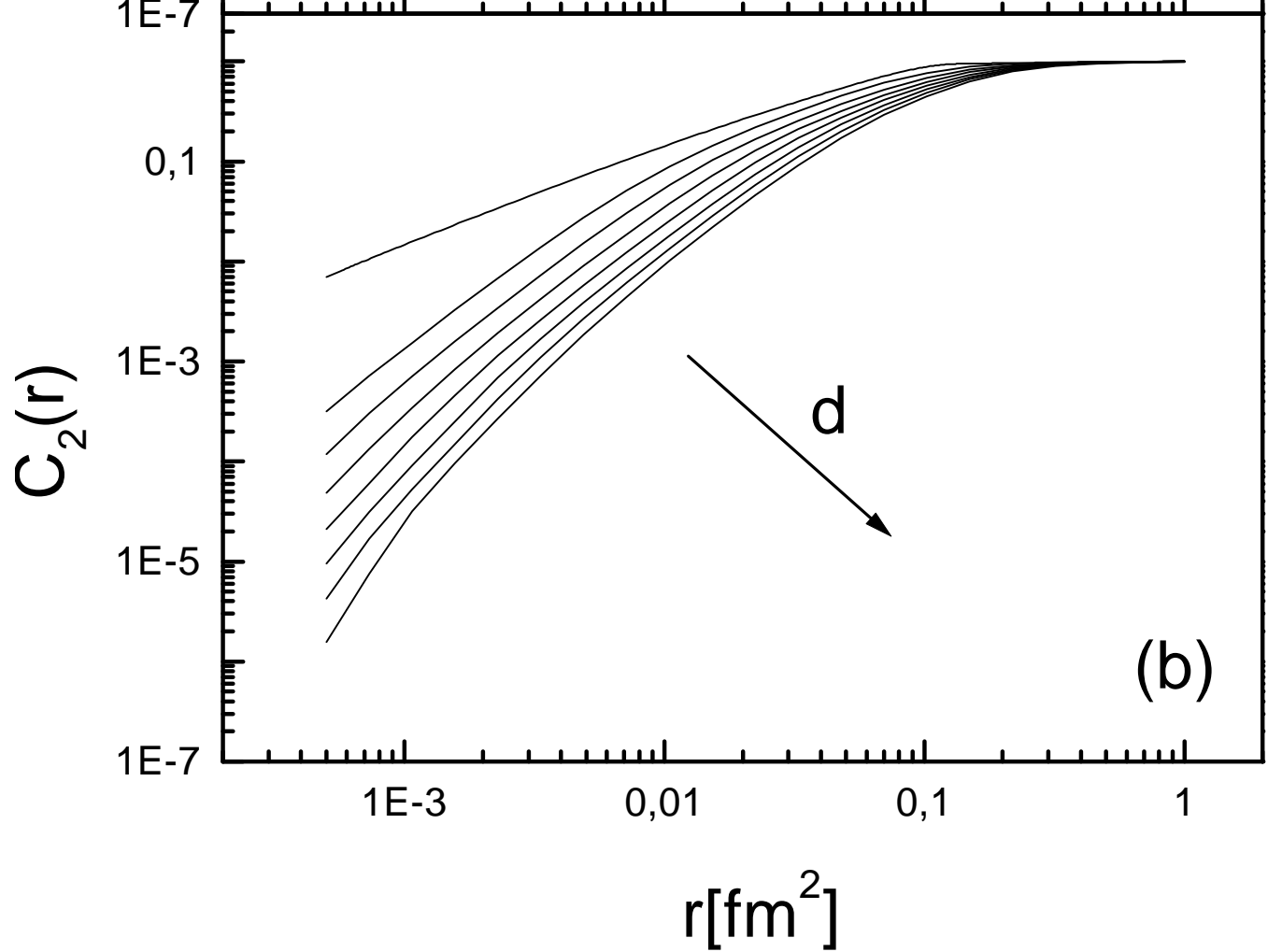
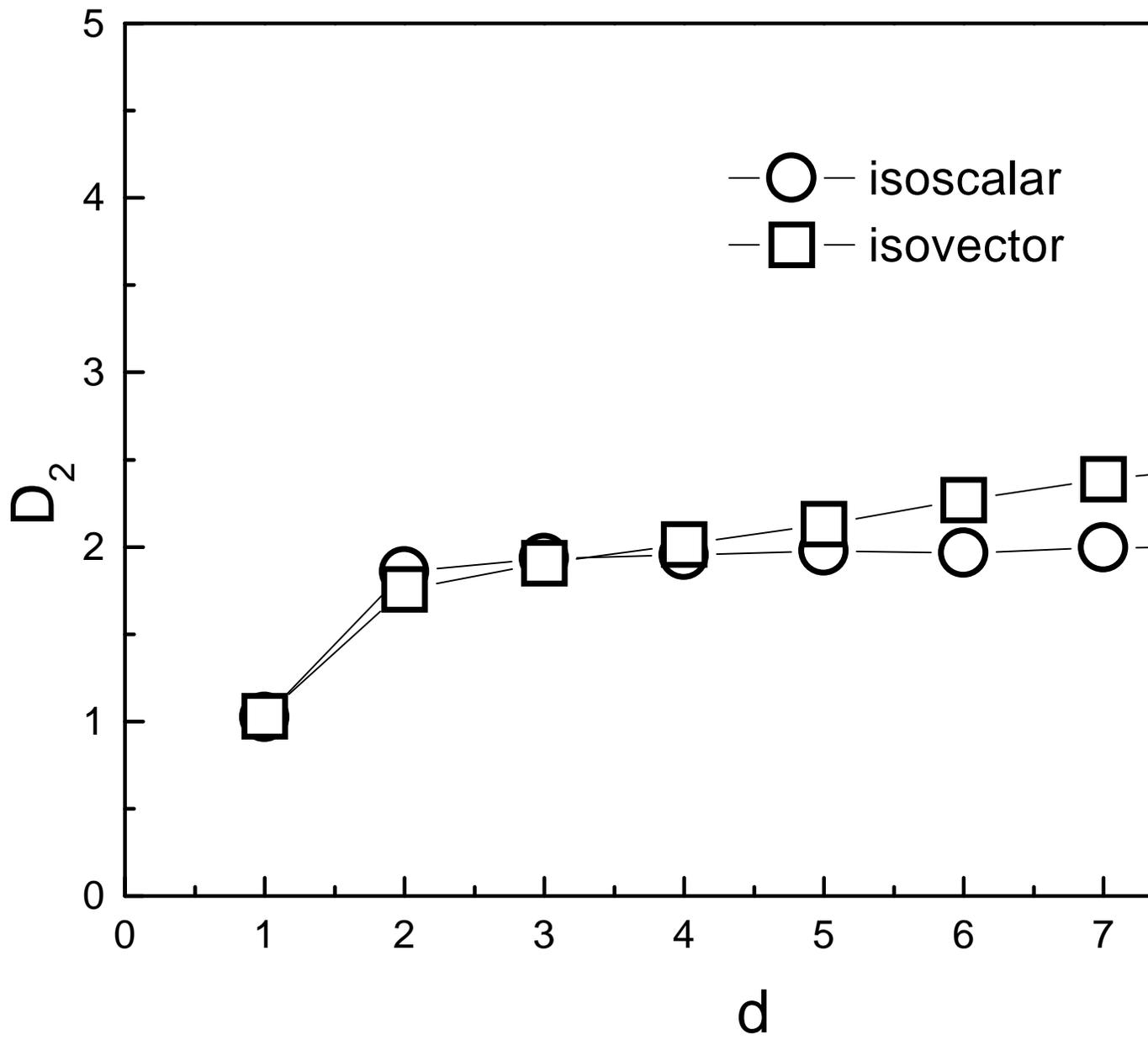
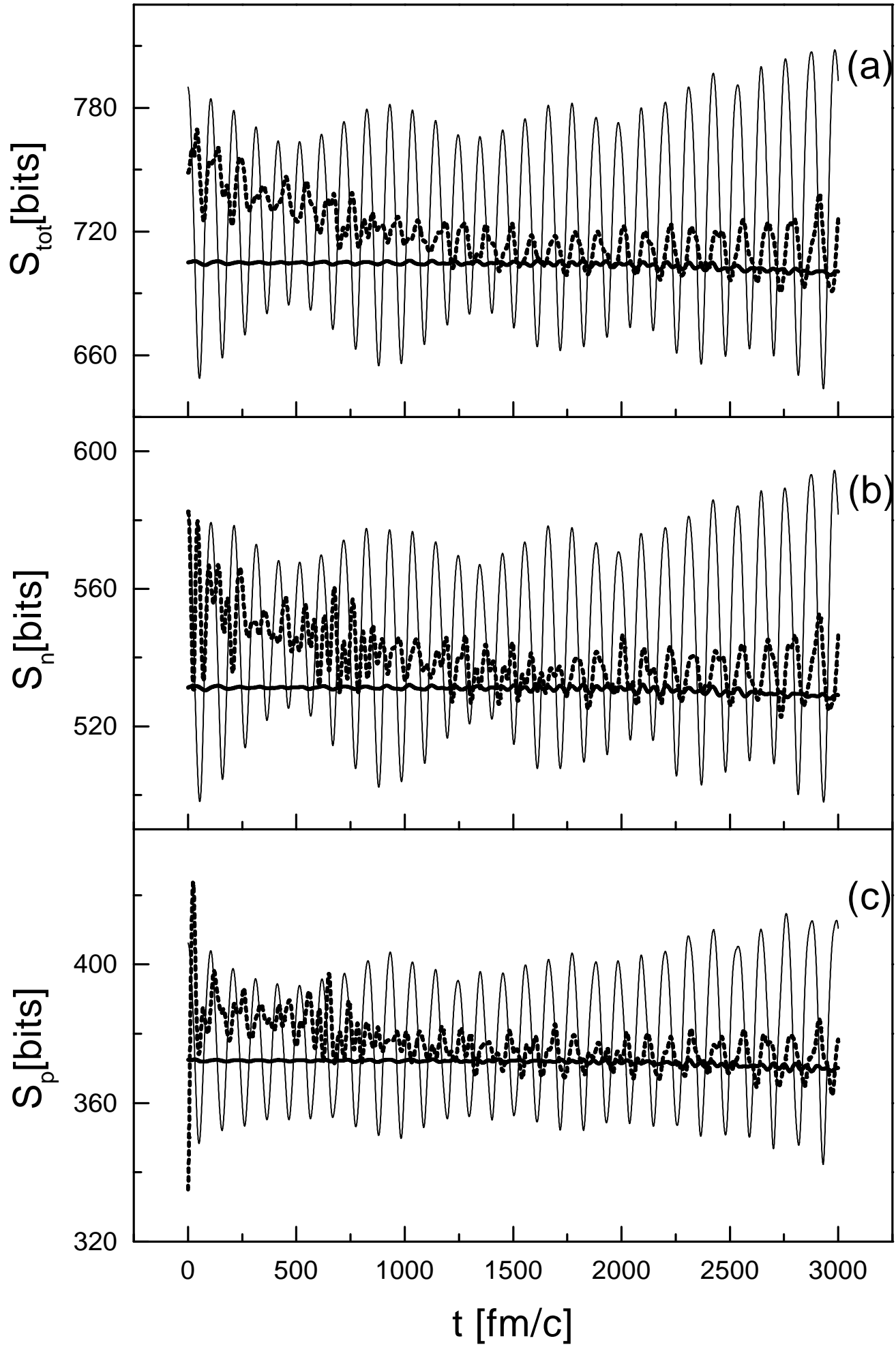
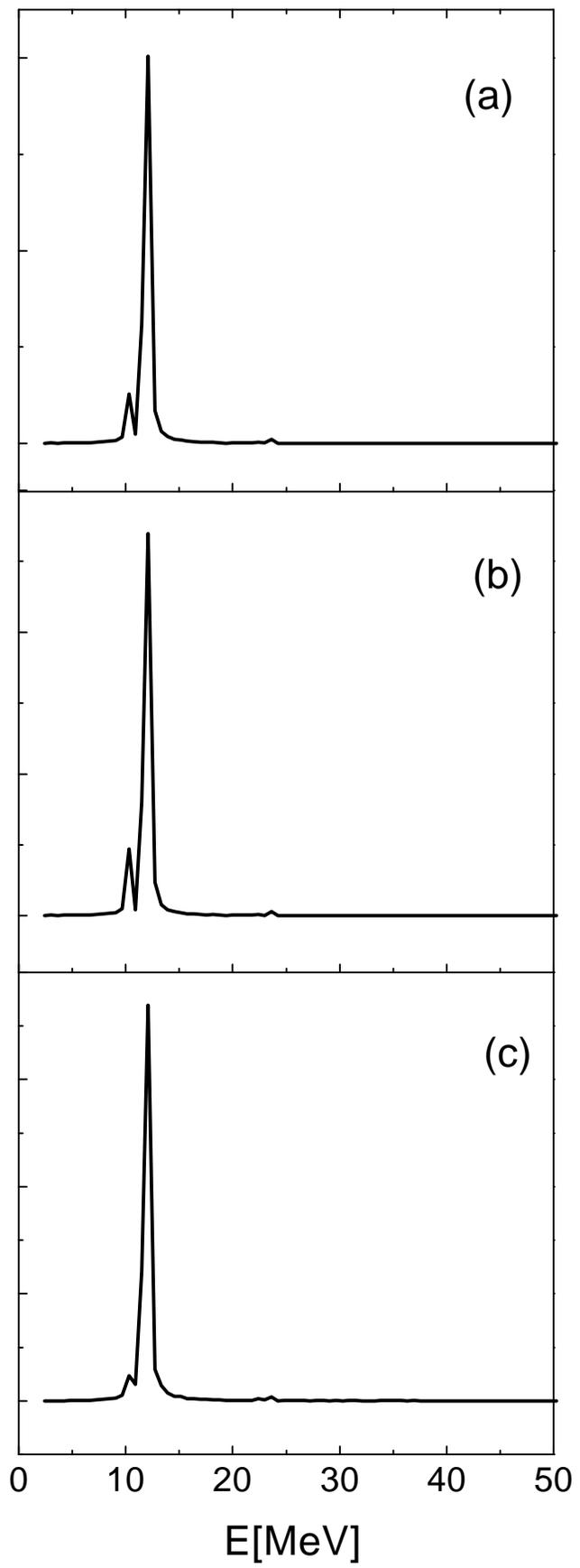
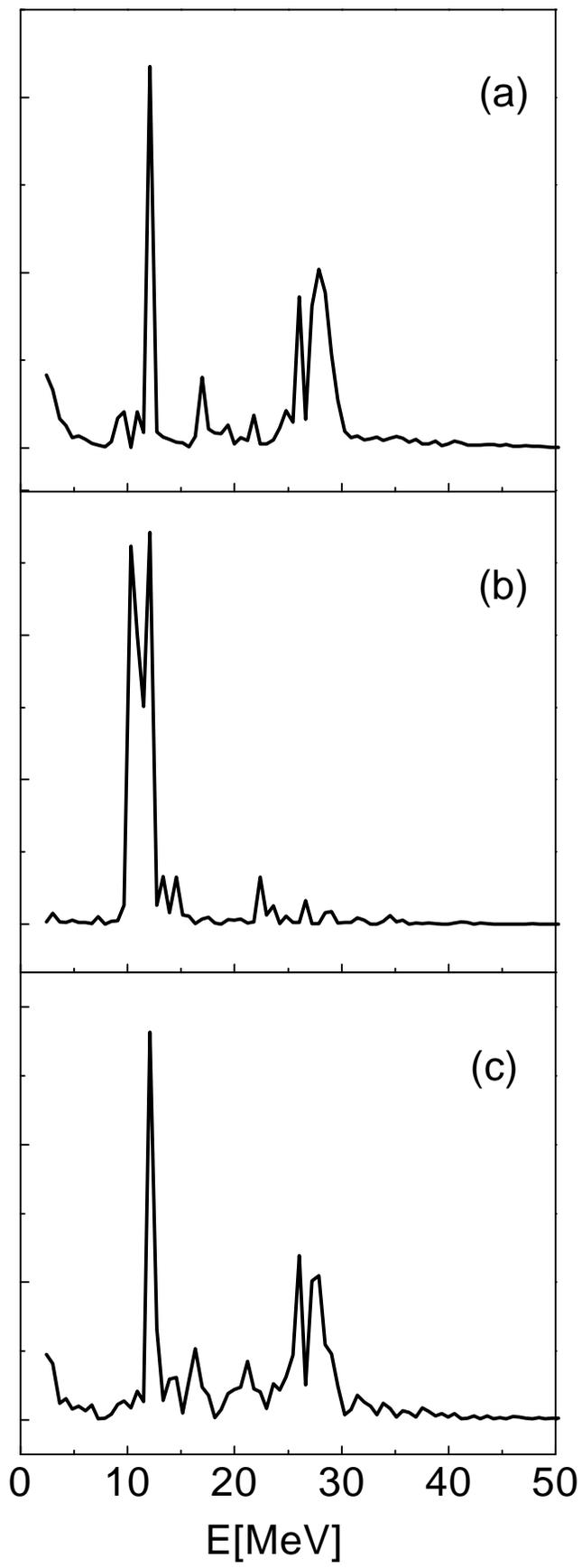
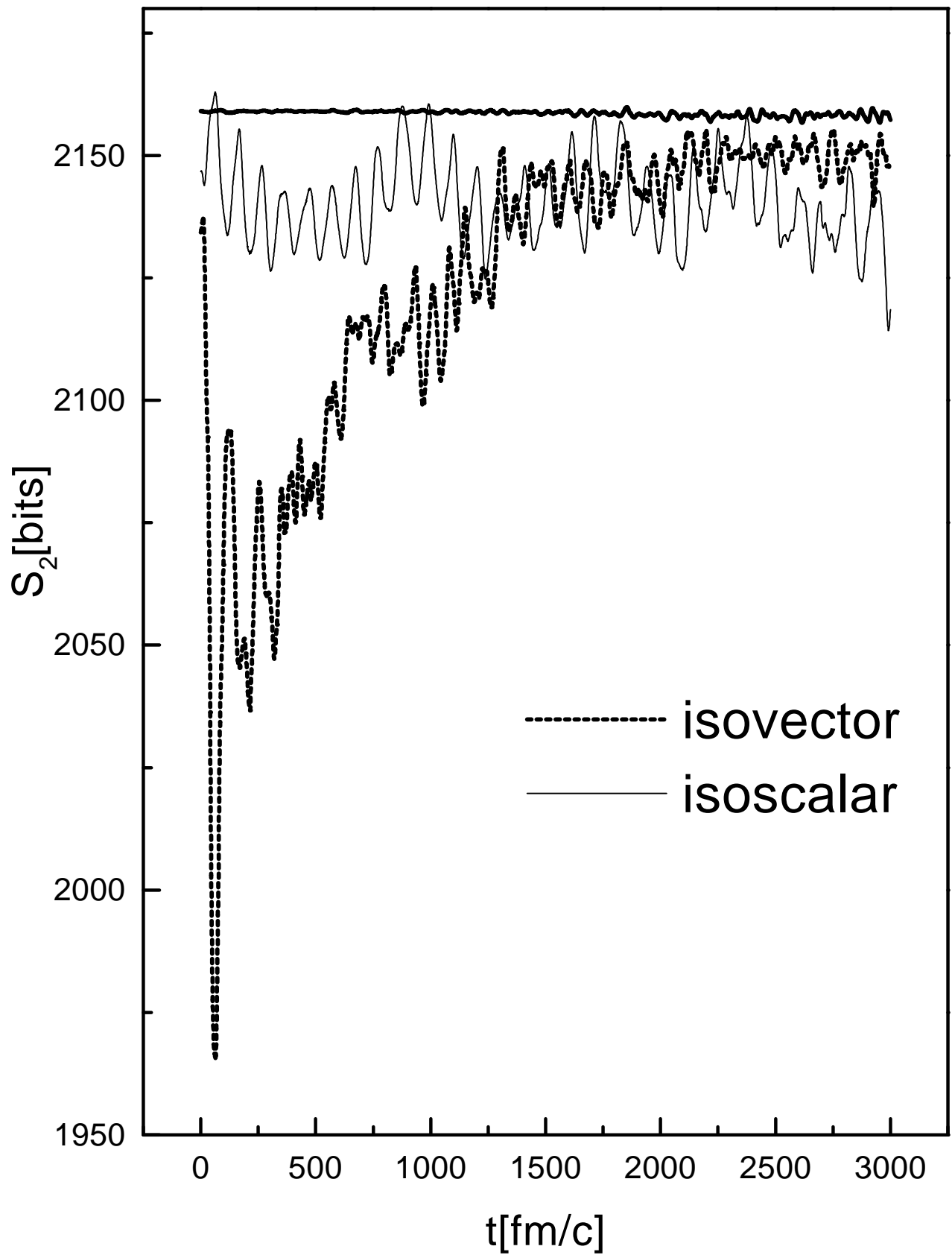
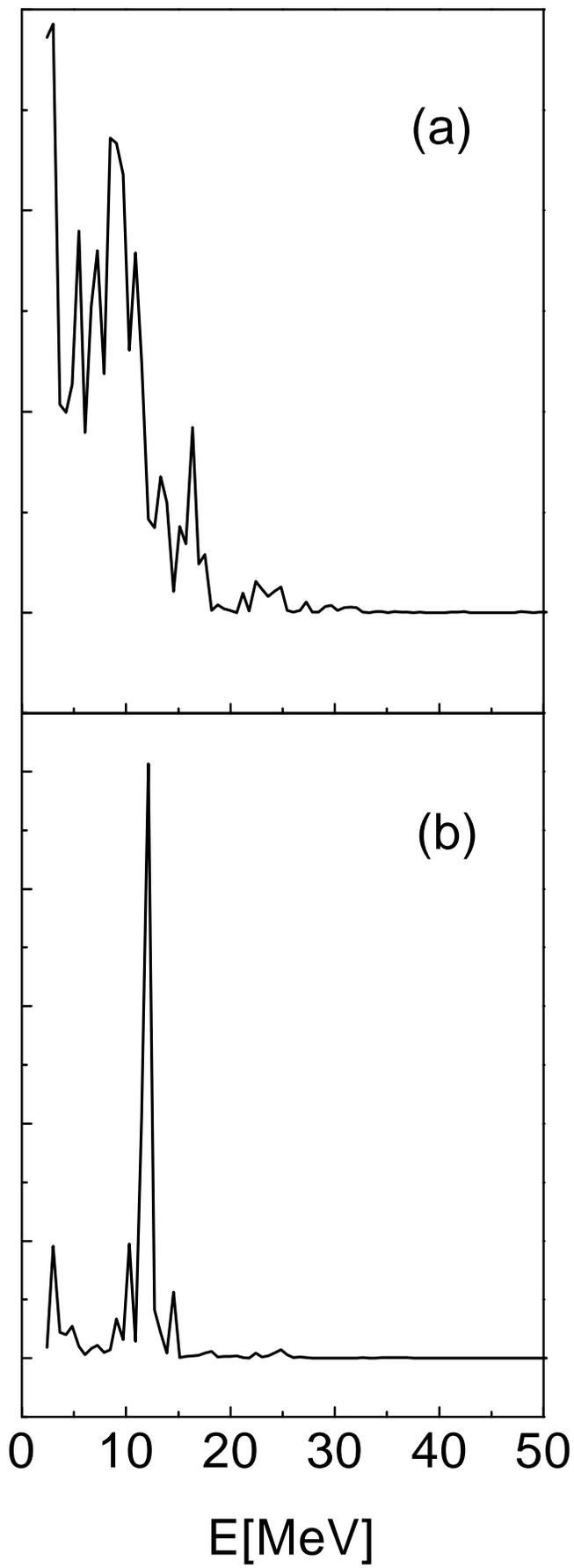
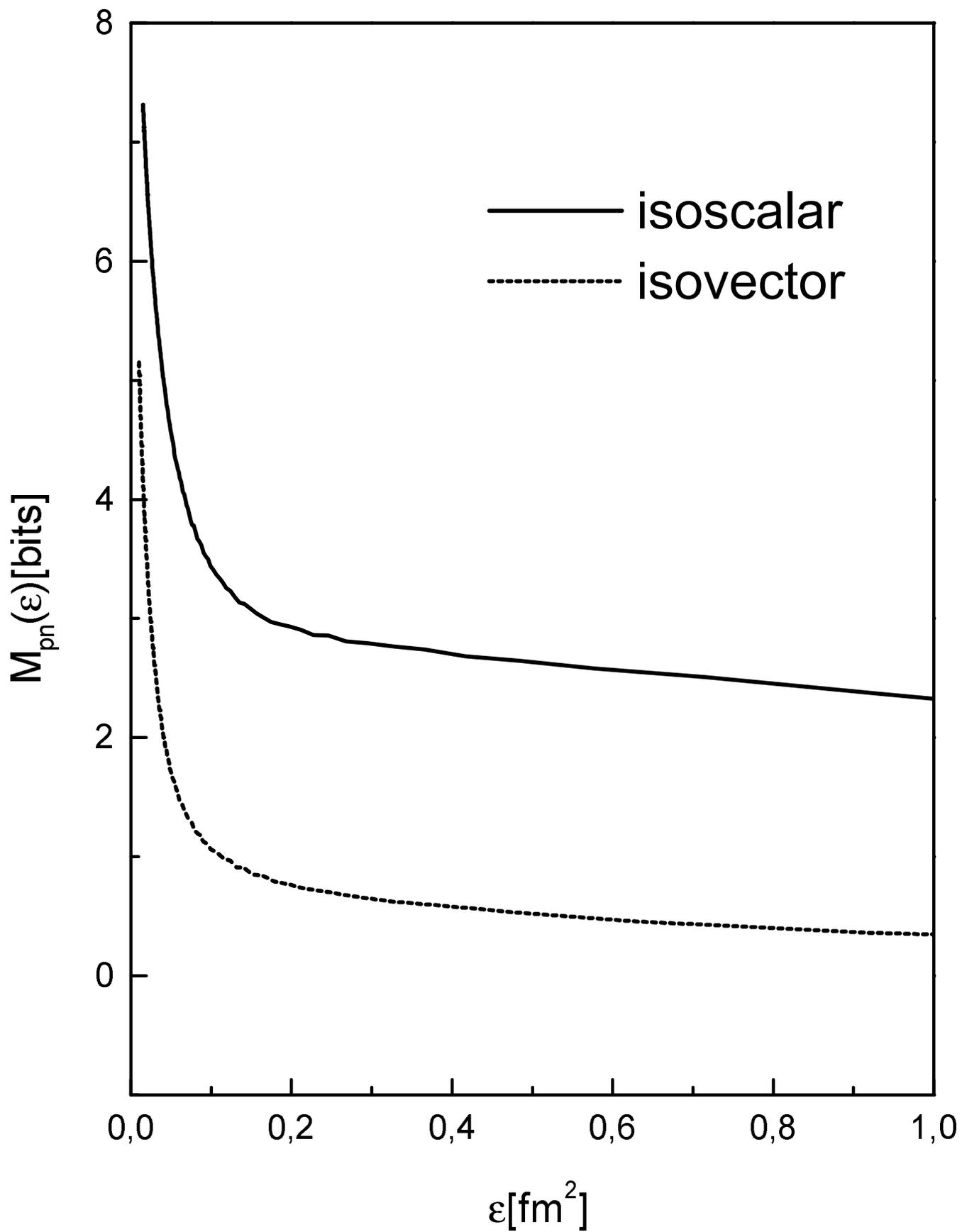
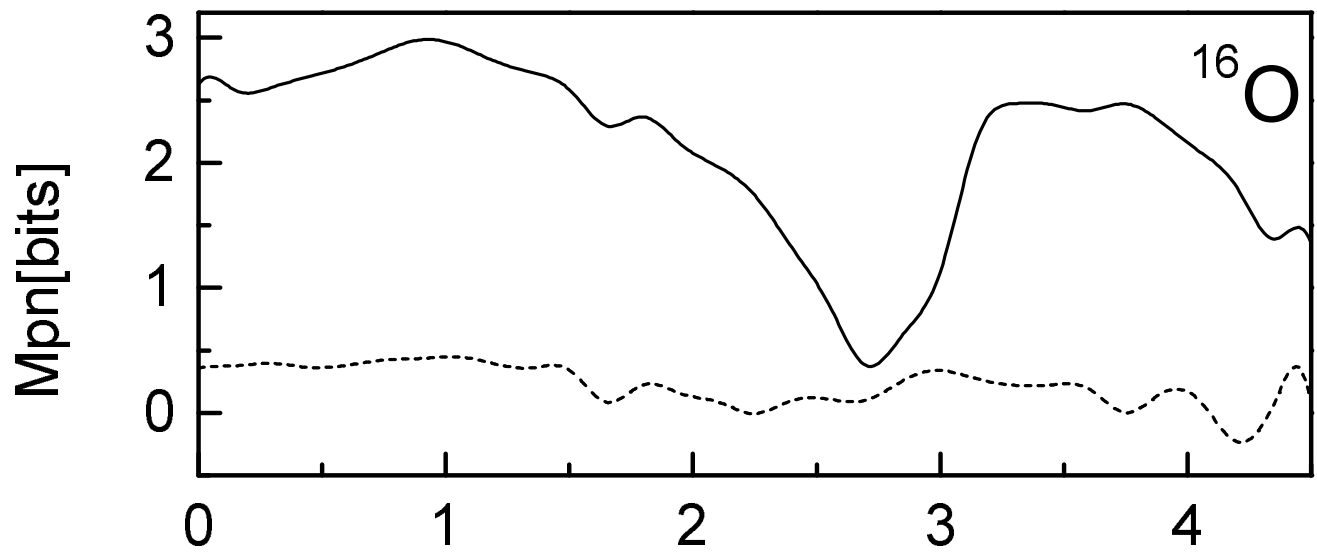
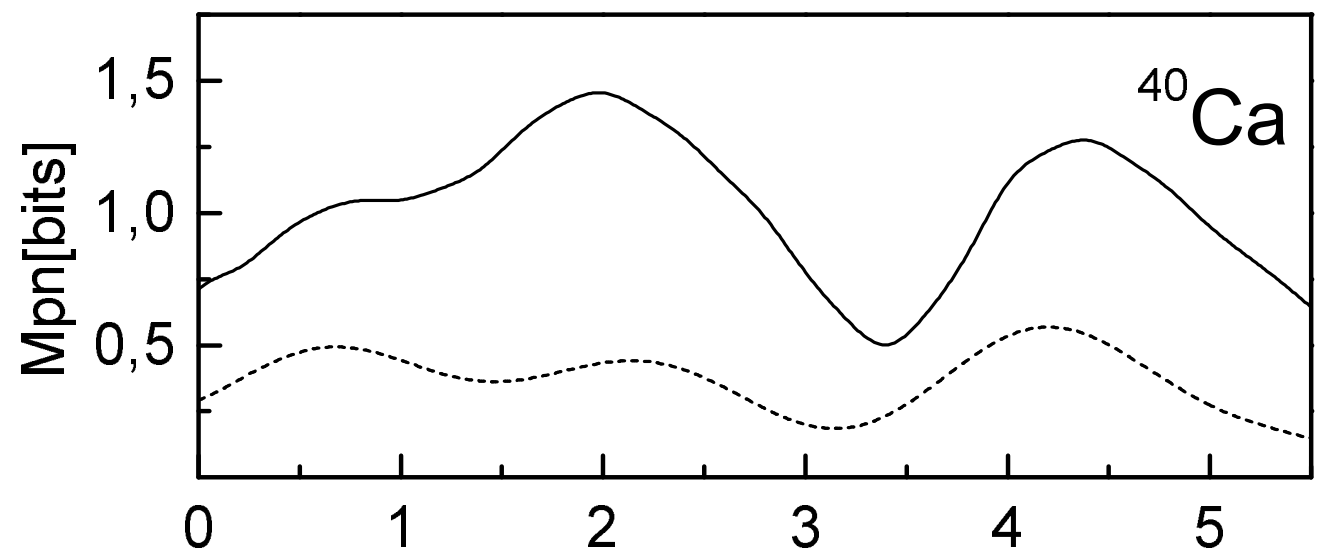
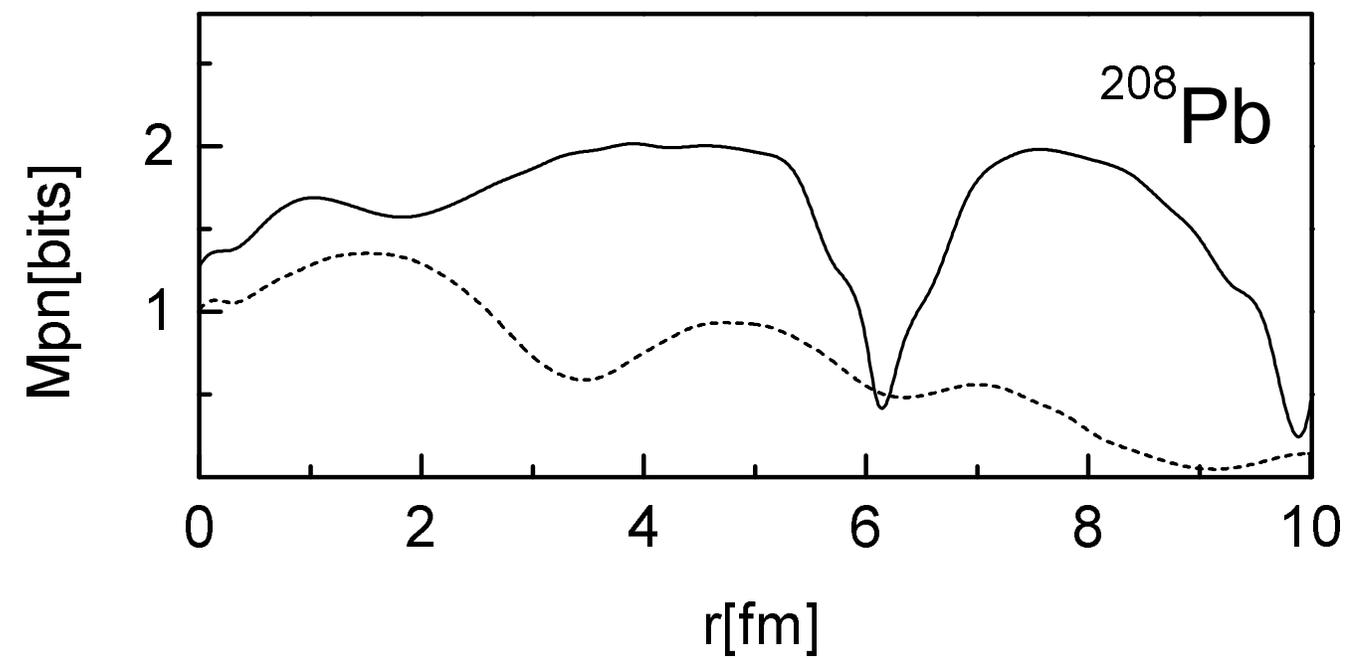
\end{document}